\PassOptionsToPackage{capitalise}{cleveref}
\PassOptionsToPackage{draft}{todonotes}
\PassOptionsToPackage{svgnames,dvipsnames}{xcolor}
\documentclass[runningheads,a4paper]{llncs}

\usepackage{caption}                \usepackage{cmll}                   \usepackage{colonequals}            \usepackage[inline]{enumitem}
\usepackage{environ}                \usepackage{etex,etoolbox}          \usepackage[nomargin,multiuser,inline,draft]{fixme} 
\fxusetheme{color}
\usepackage{graphicx}               \usepackage[utf8]{inputenc}
\usepackage{rotating}               \usepackage{todonotes}
\usepackage{url}
\usepackage{wrapfig} \usepackage{xargs}
\usepackage{xspace}

\usepackage{multirow}               \usepackage{nicefrac}
\usepackage{proof}                  \usepackage{bussproofs}
\EnableBpAbbreviations

\usepackage{subfigure}
\usepackage{thmtools, thm-restate, thm-autoref}  \usepackage{xifthen}                \usepackage{appendix}               

\usepackage{stmaryrd}
\usepackage{pdfsync}

\PassOptionsToPackage{usenames,dvipsnames,svgnames,table}{xcolor}
\usepackage{xcolor} \usepackage{tikz}                   \usetikzlibrary{shadows,arrows,shapes,automata,positioning,decorations.markings, shapes.geometric, arrows.meta, fit, calc}

\usepackage{amsmath,amssymb,amsbsy}
\usepackage{mathtools} 

\usepackage{hyperref}
\hypersetup{
  colorlinks = true,
  linkcolor = purple,
  urlcolor  = purple,
  citecolor = teal,
  anchorcolor = teal
}
\usepackage[capitalise]{cleveref}
\crefformat{enumi}{condition~#2#1#3}
\crefname{fact}{Fact}{Facts}
\Crefname{fact}{Fact}{Facts}
\crefformat{fact}{Fact~#2#1#3}

\DeclareMathSizes{10}{10}{6}{6}

\usepackage{xargs}
\usepackage[normalem]{ulem} 

\usepackage{xifthen}        \newcommand{\ifempty}[3]{\ifthenelse{\isempty{#1}}{#2}{#3}}

\DeclareGraphicsExtensions{.png,.PNG,.pdf,.PDF,.jpg,.mps,.jpeg,.jbig2,.jb2,.JPG,.JPEG,.JBIG2,.JB2
}

\newif\ifemi

\newcommandx{\preprint}[3][1=preprint,2=Springer]{
  \ifempty{#1}{}{
	 \ \\[1em]\noindent
	 \textbf{Disclaimer}
    The published version of this paper is~\cite{#1} (\copyright\ #2).
  }
}

\newcommand{\toolidcol}{red!10!blue!90}
\newcommand{\toolid}[1]{\textcolor{\toolidcol}{{\textsf{#1}}}\xspace}

\usepackage{fixme}
\fxusetheme{color}
\FXRegisterAuthor{eM}{aeM}{\color{orange} {\underline{eM}}}

\usepackage[most]{tcolorbox}
\newtcolorbox{markbox}{
  enhanced,
  breakable,
  size=minimal,
  parbox=false,
  after={\par},
  before upper={\indent},
  colback=white,
  overlay = {
	 \draw[line width=2pt]
	 (frame.north east) -| ([xshift=3mm]frame.east) |-(frame.south east);
  },
  overlay first={\draw[line width=2pt] (frame.north east) -| ([xshift=3mm]frame.south east);},
  overlay middle={\draw[line width=2pt] ([xshift=3mm]frame.north east) -- ([xshift=3mm]frame.south east);},
  overlay last={\draw[line width=2pt] ([xshift=3mm]frame.north east) |- (frame.south east);},
}

\usepackage{todonotes}
\newif\ifsubmit
\submittrue \ifsubmit
\newcommand{\eMcomm}[2][]{#2}
\newcommand{\MMcomm}[2][]{#2}
\newcommand{\ARcomm}[2][]{#2}
\newcommand{\eKcomm}[2][]{#2}
\newcommand{\hsl}[1][]{}
\else
\newcommand{\HSLtag}{\scalebox{1.25}{\begin{tikzpicture}
  \draw (0.1,0.2) -- (0.2,0.115) -- (0.3,0.2) ;
\draw (0.1,0.2) -- (0.15,0.1) ;
  \draw (0.3,0.2) -- (0.25,0.1) ;
\draw (0.12,0.1) -- (0.2,0.05) -- (0.28,0.1) ;
\draw (0.2,0) circle (0.12) ; 
  \draw (0.16,0.04) circle (0.01) ;
  \draw (0.24,0.04) circle (0.01) ;
  \draw (0.15,-0.07) -- (0.25,-0.07) ;
  \end{tikzpicture}
}}

\newcommand{\hsl}[1][]{\par
  {\color{red}\vbox{\medskip\noindent\hrulefill \\[5pt]
  \HSLtag \hspace{\stretch{1}}HIC SUNT
  LEONES \; {#1}\hspace{\stretch{1}} \HSLtag \\ \smallskip\noindent\hrulefill \\}}\par
}
\newcommand{\eMcomm}[2][check]{\ifthenelse{\equal{#1}{new}}{{\color{red}#2}}{\ifthenelse{\equal{#1}{changed}}{{\color{teal}{#2}}}{\ifthenelse{\equal{#1}{rm}}{\todo[color=black!3]{\tiny eM: removed\\'{#2}'}}{\todo[color=orange!20]{\tiny eM: \color{NavyBlue}#1}{\color{OliveGreen}{#2}}}}}}

\newcommand{\MMcomm}[2][check]{\ifthenelse{\equal{#1}{new}}{{\color{red}#2}}{\ifthenelse{\equal{#1}{changed}}{{\color{teal}{#2}}}{\ifthenelse{\equal{#1}{rm}}{\todo[color=black!3]{\tiny MM: removed\\'{#2}'}}{\todo[color=orange!20]{\tiny MM: \color{NavyBlue}#1}{\color{OliveGreen}{#2}}}}}}
\renewcommand{\MMcomm}[2][check]{\ifthenelse{\equal{#1}{new}}{{\color{red}#2}}{\ifthenelse{\equal{#1}{changed}}{{\color{teal}{#2}}}{\ifthenelse{\equal{#1}{rm}}{\todo[color=black!3]{\tiny MM: removed\\'{#2}'}}{\todo[color=orange!20]{\tiny MM: \color{NavyBlue}#1}{\color{OliveGreen}{#2}}}}}}

\newcommand{\eKcomm}[2][check]{\ifthenelse{\equal{#1}{new}}{{\color{red}#2}}{\ifthenelse{\equal{#1}{changed}}{{\color{teal}{#2}}}{\ifthenelse{\equal{#1}{rm}}{\todo[color=black!3]{\tiny EK: removed\\'{#2}'}}{\todo[color=orange!20]{\tiny EK: \color{NavyBlue}#1}{\color{Red}{#2}}}}}}

\newcommand{\ARcomm}[2][check]{\ifthenelse{\equal{#1}{new}}{{\color{red}#2}}{\ifthenelse{\equal{#1}{changed}}{{\color{teal}{#2}}}{\ifthenelse{\equal{#1}{rm}}{\todo[color=black!3]{\tiny EK: removed\\'{#2}'}}{\todo[color=orange!20]{\tiny AR: \color{NavyBlue}#1}{\color{Orange}{#2}}}}}}
\fi

\newcommand{\hidden}[1]{}
\newcommand{\hide}[1]{}

\newcommand{\cf}[2]{
  \fontsize{#1}{#1}{\selectfont{#2}}
}

\makeatletter
\newcommand{\dolist}[2]{\def\nextitem{\def\nextitem{#1}}\@for \el:=#2\do{\nextitem\textbf{\el}}}

\newcommand{\domathlist}[2]{\def\nextitem{\def\nextitem{\ensuremath{#1}}}\@for \el:=#2\do{\ensuremath{\nextitem}\textbf{\el}}}

\def\mktest#1{
  \def\transform##1+##2+##3{##1 piu' ##2 * ##3\penalty0}\do{\expandafter\transform#1}}
\def\mksubscript#1{
  \def\transform##1[##2]{##1_{##2}\penalty0}\do{\expandafter\transform#1}}
\newcommand{\mapcmd}[3][{, }]{\def\nextitem{\def\nextitem{#1}}\@for \el:=#3\do{\nextitem{#2{\el}}}}
\makeatother

\ifemi
\usepackage{showlabels}

\newcommand{\emi}[2]{
  \marginpar{\fcolorbox{red}{shadecolor}{\cf{#1}{{#2}}}}
}
\newcommand{\emic}[2]{\par
  \fcolorbox{red}{shadecolor}{\parbox{\linewidth}{ 
      \color{gray}
      \begin{description}
      \item[{\color{blue} #2}]{\sf #1}
      \end{description}}}
}
\else
\newcommand{\emi}[2]{}
\newcommand{\emic}[2]{}{}
\fi

\def\colorFun{\color{orange}}
\newcommand{\noarg}{}
\newcommand{\mkfun}[4][\colorFun]{
  \newcommand{#2}[1][#4]{{#1\ensuremath{\mathsf{#3}}}\ifempty{##1}{\noarg}{({##1})}}}

\mkfun{\head}{hd}{}
\mkfun{\tail}{tl}{}

\newcommand{\mkuop}[4][\colorFun]{
  \newcommand{#2}[1][#4]{
    {#1\textsf{#3}}
    \ifempty{##1}{}{
      \, {##1}}
  }
}

\newcommand{\conf}[1]{\ensuremath{\langle {#1} \rangle}}

\newcommand{\emptyword}{\varepsilon}

\newcommand{\qand}[1][and]{\quad\text{#1}\quad}

\newcommand{\bnfdef}{\ ::=\ }

\newcommand{\bnfmid}{\;\ \big|\ \;}

\newcommand{\eg}{\text{e.g.,}\xspace}

{\bfseries}{\rmfamily}
{\bfseries}{\rmfamily}

\makeatletter
\newcommand*{\da@rightarrow}{\mathchar"0\hexnumber@\symAMSa 4B }
\newcommand*{\da@leftarrow}{\mathchar"0\hexnumber@\symAMSa 4C }
\newcommand*{\xdashrightarrow}[2][]{\mathrel{\mathpalette{\da@xarrow{#1}{#2}{}\da@rightarrow{\,}{}}{}}}
\newcommand{\xdashleftarrow}[2][]{\mathrel{\mathpalette{\da@xarrow{#1}{#2}\da@leftarrow{}{}{\,}}{}}}
\newcommand*{\da@xarrow}[7]{\sbox0{$\ifx#7\scriptstyle\scriptscriptstyle\else\scriptstyle\fi#5#1#6\m@th$}\sbox2{$\ifx#7\scriptstyle\scriptscriptstyle\else\scriptstyle\fi#5#2#6\m@th$}\sbox4{$#7\dabar@\m@th$}\dimen@=\wd0 \ifdim\wd2 >\dimen@
    \dimen@=\wd2 \fi
  \count@=2 \def\da@bars{\dabar@\dabar@}\@whiledim\count@\wd4<\dimen@\do{\advance\count@\@ne
    \expandafter\def\expandafter\da@bars\expandafter{\da@bars
      \dabar@ 
    }}\mathrel{#3}\mathrel{\mathop{\da@bars}\limits
    \ifx\\#1\\\else
      _{\copy0}\fi
    \ifx\\#2\\\else
      ^{\copy2}\fi
  }\mathrel{#4}}
\makeatother

\newcommand{\quo}[1]{\lq\lq {#1}\rq\rq}
\def\finex{{\unskip\nobreak\hfil
\penalty50\hskip1em\null\nobreak\hfil$\diamond$
\parfillskip=0pt\finalhyphendemerits=0\endgraf}}

\definecolor{shadecolor}{rgb}{1,0.99,0.9}
\definecolor{bg}{rgb}{0.95,0.95,0.95}

\newcommand{\abcattr}[1][a]{\textsf{#1}}
\newcommand{\abccond}[1][\rho]{#1}

\def\colorExp{\color{NavyBlue}}

\newcommand{\abcexp}[1][e]{\colorExp #1}
\newcommandx{\abctuple}[1][1 = t]{\llparenthesis{#1}\rrparenthesis}
\newcommandx{\abcget}[2][1=a,2={id},usedefault=@]{
  \ptp[{#1}]{\colorOp .}\abcattr[{#2}]
}
\newcommandx{\abcptp}[2][1=a,2=\abccond,usedefault=@]{\ifempty{#1}{}{\ptp[{#1}] \ifempty{#2}{}{{\colorOp \shortmid}}} {#2}}
\newcommandx{\abcint}[6][1=a,2=\abccond,3=e,4=e',5=b,6=\abccond',usedefault=@]{
  \abcptp[{#1}][{#2}]
  \ {\colorOp \xrightarrow{\scriptstyle \abcexp[#3]\quad\abcexp[#4]}}\ 
  \abcptp[{#5}][{#6}]
}
\newcommandx{\mkabcint}[8][3=a,4=\abccond,5=e,6=e',7=b,8=\abccond',usedefault=@]{
  \node[bblock, #1] (#2) {$\abcint[{#3}][{#4}][{#5}][{#6}][{#7}][{#8}]$};
}

\tikzset{
    abccallout/.style={
      fill=green!10,
		opacity=.5,
		overlay,
		align=center,
      cloud callout,
		cloud puffs=15,
		aspect=2.5,
		cloud ignores aspect,
		cloud puff arc=100,
		shading=ball
    }
  }

\newcommandx{\abcP}[6][1=P,2=K,3=.1cm,4=1cm,5=north east,6=proc,usedefault=@]{
  \begin{tikzpicture}
	 \node[fill=blue!10, shape=circle] (#6) {$\p[#1]$};
	 \node[abccallout, above = #3 of #6, xshift=#4, callout absolute pointer={(#6.#5)}] {$#2$}
	 ;	 
	 \draw[decorate,decoration={expanding waves,angle=7,segment length = .05cm}] (#6.east) -- ++(.5cm,0)
	 ;
  \end{tikzpicture}
}

\ExplSyntaxOn
\NewDocumentCommand{\ucgreek}{m}
 {
  \str_case:nn { #1 }
   {
    {A}{\mathrm{A}}
    {B}{\mathrm{B}}
    {C}{\Sigma}
    {D}{\Delta}
    {E}{\mathrm{E}}
    {F}{\Phi}
    {G}{\Gamma}
    {H}{\mathrm{H}}
    {I}{\mathrm{I}}
    {J}{\Theta}
    {K}{\mathrm{K}}
    {L}{\Lambda}
    {M}{\mathrm{M}}
    {N}{\mathrm{N}}
    {O}{\mathrm{O}}
    {P}{\Pi}
    {Q}{\mathrm{X}}
    {R}{\mathrm{P}}
    {S}{\Sigma}
    {T}{\mathrm{T}}
    {U}{\Upsilon}
{W}{\Omega}
    {X}{\Xi}
    {Y}{\Psi}
    {Z}{\mathrm{Z}}
   }
 }
\NewDocumentCommand{\lcgreek}{m}
 {
  \str_case:nn { #1 }
   {
    {a}{\alpha}
    {b}{\beta}
    {c}{\varsigma}
    {d}{\delta}
    {e}{\varepsilon}
    {f}{\varphi}
    {g}{\gamma}
    {h}{\eta}
    {i}{\iota}
    {j}{\vartheta}
    {k}{\kappa}
    {l}{\lambda}
    {m}{\mu}
    {n}{\nu}
    {o}{o}
    {p}{\pi}
    {q}{\chi}
    {r}{\rho}
    {s}{\sigma}
    {t}{\tau}
    {u}{\upsilon}
{w}{\omega}
    {x}{\xi}
    {y}{\psi}
    {z}{\zeta}
   }
 }
\ExplSyntaxOff

 \usepackage{listings}
\usepackage{mfirstuc}
\usepackage{xstring}

\newcommand{\gsubs}[2]{^{#1} / _{#2}}
\newcommandx{\gsubst}[3][1=\aM,2=q,3=q',usedefault=@]{
  \left \{\gsubs{#3}{#2} \right \}#1
}
\newcommandx{\gsubsts}[5][1=\aM,2=q,3=q',4=q,5=q',usedefault=@]{
  \left \{\gsubs{#3}{#2}, \gsubs{#5}{#4} \right \}#1
}

\newcommandx{\isgc}[2][1=/tmp/__gc__]{
  \makeatletter\def\@gcpar{-o "#1" -gc "#2"}\makeatother
  \immediate\write18{gc2latex \@gcpar}
  \input{#1.tex}
}

\newcommandx{\isgcfile}[2][1=/tmp/__gc__]{
  \makeatletter\def\@gcpar{-o "#1" "#2"}\makeatother
  \immediate\write18{gc2latex \@gcpar}
  \input{#1.tex}
}

\newif\ifcp
\cpfalse
\newcommand{\gname}[1][i]{\ifcp{\colorNode{\scriptstyle\textsf{#1}}}\else{}\fi}

\newif\ifguard
\guardfalse
\newcommand{\aguard}{\ifguard{\colorGuard \phi}\else{}\fi}

\def\colorGuard{\color{cyan}}
\def\colorPtp{\color{blue}}

\def\colorFun{\color{Navy}}

\def\colorOp{\color{OliveGreen}}
\def\colorNode{\color{LightCoral}}
\def\colorR{\color{OliveGreen}}
\def\colorE{\color{orange}}

\def\colorMsg{\color{BrickRed}}

\newcommand{\fillcolor}{orange!5}

\newcommandx{\wellpar}[2][1={\aG},2={\aG'},usedefault=@]{
  \mathit{wf}({#1}, {#2})
}

\newcommand{\amsg}[1][m]{\ensuremath{\mathsf{\colorMsg{#1}}}\xspace}
\newcommand{\msg}[1][m]{\amsg[{#1}]}
\newcommand{\ptp}[1][A]{\ensuremath{\mathsf{\colorPtp{\capitalisewords{#1}}}}\xspace} 
\newcommand{\p}{\ptp[a]\xspace}
\newcommand{\q}{{\ptp[B]}\xspace}

\newcommand{\sndint}[1][{\gint[]}]{\mathrm{snd}(#1)}
\newcommand{\rcvint}[1][{\gint[]}]{\mathrm{rcv}(#1)}
\newcommandx{\ggcommon}[3][1=\ptp,2={\aH},3={\aH'},usedefault=@]{f_{#1}}
\newcommandx{\opair}[2][1={\ae},2={\ae'},usedefault=@]{\conf{{#1},{#2}}}
\newcommandx{\hopair}[2][1={\aE},2={\aE'},usedefault=@]{\llparenthesis\, {#1},{#2}\, \rrparenthesis}
\newcommandx{\wf}[2][1={\aG},2={\aG'},usedefault=@]{wf({#1}, {#2})}
\newcommandx{\wb}[2][1={\aG},2={\aG'},usedefault=@]{wb({#1}, {#2})}
\newcommandx{\ws}[2][1={\aG},2={\aG'},usedefault=@]{ws({#1}, {#2})}

\newcommandx{\widx}[2][1={\aW},2={i},usedefault=@]{{#1}[{#2}]}
\newcommandx{\outop}[2][1=\gname,2={}]{{\colorOp{!}}^{{#1}{#2}}}
\newcommandx{\inop}[2][1=\gname,2={}]{{\colorOp{?}}^{{#1}{#2}}}
\newcommandx{\aout}[5][1=a,2=b,3={},4=m,5={},usedefault=@]{
  \achan[#1][#2] \outop[{#3}] {\msg[#4]} {#5}
}
\newcommandx{\ain}[5][1={\p},2={\q},3=\gname,4=m,5={},usedefault=@]{
  \achan[#1][#2] \inop[{#3}] {\msg[{#4}]}{#5}
}
\newcommandx{\adep}[1][1={}]{
  \conf{ \aout[@][@][@][@][{#1}], \ain[@][@][@][@][{#1}]}
}

\newcommandx{\hproj}[2][1=\aH, 2=\ptp, usedefault=@]{
  \ifempty{#1}{}{{#1}}\ifempty{#2}{}{{^{\scriptscriptstyle @{#2}}}}
}
\newcommandx{\eproj}[2][1=\aE,2=A, usedefault=@]{
  {{#1}}\ifempty{#2}{}{{^{\scriptscriptstyle @{{\ptp[#2]}}}}}
}

\tikzset{
  component/.style={
    draw,
    fill = white,
    minimum width = 1.5cm,
    minimum height = .5cm,
    drop shadow
  }
}
\tikzset{
  file/.style={
    thin,
    fill = blue!5,
    font = \tt\scriptsize,
text width = .8cm,
    minimum width = 1.0cm,
    minimum height = .5cm,
    drop shadow
  }
}

\tikzset{
  dataflow/.style={
    thick,
    draw, ->, >=latex,
    dashed,
    OliveGreen
  }
}

\tikzset{
  pipeline/.style={
    thick,
    draw, ->, >=latex,
    double,
    red
  }
  
}

\tikzset{
  pomsetcloud/.style={
    cloud,
	 cloud puffs=20,
	 cloud ignores aspect,
	 minimum height=.1cm,
	 minimum width=2cm,
	 fill=blue!10,
	 opacity=.5,
	 draw
  }
}

\newcommand{\apom}{r}

\newcommand{\aR}[1][R]{{\colorR{#1}}}

\newcommandx{\detM}[1][1=\aCM,usedefault=@]{\Delta({#1})}
\newcommandx{\testsys}[2][1=\aM,2=\atestcase,usedefault=@]{#1 \otimes  #2}
\newcommand{\atestcase}[1][T]{#1}

\newcommandx{\aCFSM}[3][1=\aQ,2=\aQzero,3=\aTrs]{(#1,#2,#3)}

\newcommandx{\aQfinal}[1][1=,usedefault=@]{
  {\ifempty{#1}{F}{F_{#1}}}
}

\tikzset{
mycfsm/.style={
        font=\footnotesize,
        initial where=left,
        ->,>=stealth,auto,
scale=1, every state/.style={
			 inner sep=.5pt,
			 minimum size = 4pt,
			 transform shape
		  },
        every edge/.style = {carrow},
        baseline=(current bounding  box.center),
        initial text={}
  },
  cfsm/.style={
         node distance=2.2cm and 1cm,
         scale=.85,
         transform shape,
         smooth,
         every state/.style = {cnode},
         every edge/.style = {carrow}
  },
  cnode/.style={
    shape=circle,
    minimum size = 0mm,
    inner sep = 1pt,
    font=\tiny,
    draw
  },
  carrow/.style={
    ->,
    shorten >=1pt,
    >=stealth',
    auto,
    font=\scriptsize,
    draw,
    sloped
  }
}

\newcommandx{\synccfsmanimation}[3][1=a,2=c,3=b,usedefault=@]{
  \begin{overlayarea}{\textwidth}{\textheight}\vspace{-.5cm}
  \begin{block}{Internal step: $\aCS \trans \emptyword \aCS'$}\centering
	 \begin{tikzpicture}[mycfsm]
		\node[machinecloud, initial, initial text={\ptp[{#1}]}, initial distance = .3cm] (a) {$\bullet\phantom{\trans{\aout[#1][#2][]} \circ}$};
		\node[right = of a](dots1){$\cdots$};
		\node<1>[machinecloud, fill=cyan!10, initial, initial text={\ptp[{#2}]}, right = of dots1, initial distance = .3cm] (c) {$\bullet \trans \emptyword \circ$};
		\node<2->[machinecloud, initial, initial text={\ptp[{#2}]}, right = of dots1, initial distance = .3cm] (c) {$\circ \trans \emptyword \textcolor{orange}{\bullet}$};
		\node[right = of c](dots2){$\cdots$};
		\node[machinecloud, initial, initial text={\ptp[{#3}]}, right = of dots2, initial distance = .3cm] (b) {$\bullet\phantom{\trans{\ain[#1][#2][]} \circ}$};
	 \end{tikzpicture}
  \end{block}
  \begin{block}<3->{Interaction: $\aCS \trans {\gint[][#1][@][#3]}  \aCS'$}\centering
	 \begin{tikzpicture}[mycfsm]
		\node<3>[machinecloud, fill=cyan!10, initial, initial text={#1}, initial distance = .3cm] (a) {$\bullet \trans{\aout[#1][#3][]} \circ$};
		\node<4->[machinecloud, initial, initial text={#1}, initial distance = .3cm] (a) {$\circ \trans \aout \textcolor{orange}{\bullet}$};
		\node[right = of a](dots1){$\cdots$};
		\node[machinecloud, initial, initial text={#2}, right = of dots1, initial distance = .3cm] (c) {$\bullet$\phantom{$\trans \emptyword \circ$}};
		\node[right = of c](dots2){$\cdots$};
		\node<3>[machinecloud, fill=cyan!10, initial, initial text={#3}, right = of dots2, initial distance = .3cm] (b) {$\bullet \trans{\ain[#1][#3][]} \circ$};
		\node<4->[machinecloud, initial, initial text={#3}, right = of dots2, initial distance = .3cm] (b) {$\circ \trans{\ain[#1][#3][]} \textcolor{orange}{\bullet}$};
	 \end{tikzpicture}
  \end{block}
\end{overlayarea}
  \transdissolve<2>
  \transdissolve<3>
  \transdissolve<4>
}

\newcommandx{\choranimation}[5][1=1,2=1,3=,4=.7,5=,usedefault=@]{
\begin{overlayarea}{#1\textwidth}{#2\textheight}
    \begin{tikzpicture}[
      node distance=2cm and 1cm,
      scale=#4,
      every node/.style={transform shape},
]
      \node [choreo] (global){Choreography $\aG$ \\ global viewpoint};
\node [choreo, below =of global] (type1) {$\aCM_1$ \\ Local viewpoint$_1$};
      \node [choreo, right =of type1] (typei) {$\aCM_i$ \\ Local viewpoint$_i$};
      \node [choreo, right =of typei] (typen) {$\aCM_n$ \\ Local viewpoint$_n$};
\node<+-> (synctxt) [left = of global] {\ifempty{#5}{\textcolor{Navy}{\bf Synchrony}}{}};
      \node<.-> (asynctxt) [left = of type1] {\ifempty{#5}{\textcolor{Navy}{\bf Asynchrony}}{}};
		\ifempty{#5}{\node<+-> (wf) [right = of typen, fill=blue!10, drop shadow] {spec,no code};}
		\ifempty{#5}{\node<+-> (nocode) [above right = of typen, fill=blue!10, drop shadow] {Well-formedness};}
\path<+-> [bigar] (global) edge[sloped,above] node {\color{OliveGreen}Project} (type1);
      \path<.-> [bigar] (global) edge[sloped,above] node {\color{OliveGreen}Project} (typei);
      \path<.-> [bigar] (global) edge[sloped,above] node {\color{OliveGreen}Project} (typen);
\path[elli] (type1) -- (typei);
      \path[elli] (typei) -- (typen);
\node<+-> [process, below=of type1] (proc1) {Component$_1$};
      \node<.-> [process, below=of typei] (proci) {Component$_i$};
      \node<.-> [process, below=of typen] (procn) {Component$_n$};
\path<.-> [bigar,->,dashed,gray] (type1) edge[sloped,above] node {Verify} (proc1);
      \path<.-> [bigar,->,dashed,gray] (typei) edge[sloped,above] node {Verify} (proci);
      \path<.-> [bigar,->,dashed,gray] (typen) edge[sloped,above] node {Verify} (procn);
\path<.-> [elli] (proc1) -- (proci);
      \path<.-> [elli] (proci) -- (procn);
		\ifempty{#3}{}{
\node<+-> [process, right=of procn,xshift=4cm] (evolve1) {Component'$_1$};
		  \node<.-> [process, right=of evolve1] (evolvei) {Component'$_i$};
		  \node<.-> [process, right=of evolvei] (evolven) {Component'$_n$};
\path<.-> [bigar,blue,dotted] (procn) edge node [above] {evolve/replace/compose} (evolve1);      
        \path<.-> [elli] (evolve1) -- (evolvei);
        \path<.-> [elli] (evolvei) -- (evolven);
\node<+-> [choreo, above=of evolve1, yshift=1.5cm] (t11) {New $\aCM'_1$ \\ Local viewpoint$_1$};
		  \node<.-> [choreo, above=of evolvei, yshift=1.5cm] (t1i) {New $\aCM'_i$ \\ Local viewpoint$_i$};
		  \node<.-> [choreo, above=of evolven, yshift=1.5cm] (t1n) {New $\aCM'_n$ \\ Local viewpoint$_n$};
\path<.-> [elli] (t11) -- (t1i);
        \path<.-> [elli] (t1i) -- (t1n);
\path<.-> [bigar,->,dashed,gray] (evolve1) edge[sloped,above] node {Extract} (t11);
        \path<.-> [bigar,->,dashed,gray] (evolvei) edge[sloped,above] node {Extract} (t1i);
        \path<.-> [bigar,->,dashed,gray] (evolven) edge[sloped,above] node {Extract} (t1n);
\node<+> [above=of t1i, yshift=.5cm] (qm) {\Huge \textcolor{red}{ ??? }};
		  \node<.-> [above=of t1i] (dummy) {};
		  \node<+-> [choreo,above=of dummy, yshift=-.5cm] (global') {New choreography $\aG'$ \\ global viewpoint};
\path<.-> [bigar,-] (t11) -- (dummy);
        \path<.-> [bigar,->] (t1i) edge node[right,xshift=1em] {\color{OliveGreen}Synthesise} (global');
        \path<.-> [bigar,-] (t1n) -- (dummy);
		}
    \end{tikzpicture}
  \end{overlayarea}
}

\newcommandx{\cm}[2][1=\ptp, 2=\aM]{{#2}_{#1}}
\newcommandx{\achan}[2][1=A,2=B,usedefault=@]{{\ptp[#1]}{\,}{\ptp[#2]}}

\newcommand{\oact}{\outop[]}
\newcommand{\iact}{\inop[]}
\newcommand{\tset}{\to}

\newcommand{\trans}[2][{}]{\,\xrightarrow{#2}_{#1}\,}

\newcommandx{\acfsmout}[3][1=A,2=B,3=m,usedefault=@]{\achan[{#1}][{#2}] \oact {\msg[{#3}]}}
\newcommandx{\acfsmin}[3][1=A,2=B,3=m,usedefault=@]{\achan[{#1}][{#2}] \iact {\msg[{#3}]}}
\newcommandx{\fsaout}[2][1={\p},2={},usedefault=@]{
  \ptp[#1] \ \outop[]\ \msg[{#2}]
}
\newcommandx{\fsain}[2][1={\p},2={},usedefault=@]{
  \ptp[#1] \ \inop[]\ \msg[{#2}]
}

\makeatletter
\newcommand{\linenumfontsize}{\@setfontsize{\linenumfontsize}{3pt}{3pt}}
\makeatother
\lstdefinelanguage{sys}{
	commentstyle=\color{Gray},
morecomment=[s]{[}{]},
keywords=[0]{system,of,do,end},	keywordstyle=\color{orange}\bfseries,
}

\newcommand{\bracketColor}[1]{\textcolor{cyan!50!blue!80}{#1}}
\lstdefinelanguage{sgc}{
  commentstyle=\color{Gray},
  morecomment=[l]{..},
  morecomment=[n]{[[}{]]},
  keywords=[0]{repeat,branch,sel,let,in,with,unless},
  morekeywords=[1]{test},
  keywordstyle=[0]\color{teal}\sffamily,
  keywordstyle=[1]\color{red}\sffamily,
  mathescape=true,
  literate=
  {=}{{$\colorOp{=}$}}1
  {->}{{${\colorOp \xrightarrow{}}\ $}}2
  {=>}{{${\colorOp \Rightarrow{}}$}}2
  {:}{{$\colorOp{\colon}$}}1
  {|}{{$\gparop$}}1
  {;}{{$\gseqop$}}1
  {+}{{$\gchoop$}}1
  {(}{{\bracketColor{(}}}1
  {)}{{\bracketColor{)}}}1
  {\{}{{\bracketColor{\{}}}1
  {\}}{{\bracketColor{\}}}}1
  {[}{{\bracketColor{[}}}1
  {]}{{\bracketColor{]}}}1
  {[[}{{\color{Gray}{[[}}}2
  {]]}{{\color{Gray}{]]}}}2
  {(o)}{{\gempty}}1
}

\newcommand{\aG}{\mathsf{G}}

\newcommand{\gseqop}{{\colorOp ;}\,}
\newcommand{\gparop}{{\colorOp \ \ensuremath{\mid}\ }}
\newcommand{\gchoop}{{\colorOp \ +\ }}
\newcommand{\grecop}{{\colorOp *}}
\newcommand{\grecopp}{{\colorOp{@}}}
\newcommandx{\nmerge}[2][1={i},2={},usedefault=@]{
  \ifempty{#2}{
    \ifempty{#1}{\mu}{\gname[-{#1}]}
  }{-{#2}}
}

\mkfun{\esbj}{sbj}{\ae}

\makeatletter \@ifclassloaded{exam-paper}{}{\makeatletter \@ifclassloaded{test}{}\makeatother }
\makeatother

\newcommandx{\gnode}[2][1=i,2=\gint,usedefault=@]{
  \ifcp{
    \ifempty{#1}{#2}{\gname[#1].\big({#2}\big)}
  }
  \else
  {#2}
  \fi
}

\makeatletter
\def\mklogtype#1{
  \def\nextitem{\def\nextitem{\logcat\penalty0}}\@for\el:=#1\do{\nextitem\aeventtype[][{\el}]}}

\def\mklog#1{
  \ifempty{#1}{\emptylog}{
    \def\nextitem{\def\nextitem{\logcat\penalty0}}\@for\el:=#1\do{\nextitem\aevent[][{\el}]}}
}

\def\mksys#1{
  \def\transform##1[##2]{(\afish[][##1],\mklog{##2})\parop\penalty0}\@for\el:=#1\do{\expandafter\transform\el}}
\makeatother

\newcommand{\gempty}{\ensuremath{\circledcirc}}\newcommandx{\refgint}[3][1=A,2=\msg,3=B,usedefault=@]{
  \ptp[#1] {\colorOp \xdashrightarrow[{}]{\msg[{#2}]}}{
    \renewcommand{\do}[1]{\ptp[##1]}
	 \docsvlist{#3}
  }
}
\newcommandx{\gint}[4][1=i,2=A,3=m,4=B,usedefault=@]{
  \scalebox{.8}{$
	 \ptp[#2] {\colorOp \xrightarrow{\scriptstyle \gname[#1]}} \ptp[#4] {\colorOp \colon} {\msg[{#3}]}
   $}
}
\newcommandx{\gout}[4][1=\gname,2=\ptp,3=m,4={\ptp[C]},usedefault=@]{
  \achan[{#2}][{#4}] {\colorOp {\colorOp{!}}} {\msg[{#3}]}
}
\newcommandx{\gin}[4][1=\gname,2=\ptp,3=m,4={\ptp[C]},usedefault=@]{
  \achan[{#2}][{#4}] {\colorOp{?}} {\msg[{#3}]}
}
\newcommandx{\gseq}[3][1=\gname,2={\aG},3={\aG'},usedefault=@]{
  \def\ggraph{{#2} \gseqop {#3}}
  \ggraph
}
\newcommand{\ginfix}[4]{
  \def\ggraph{{#2} {#4} {#3}}
  \gnode[#1][\ggraph]
}
\newcommandx{\gpar}[3][1=i,2={\aG},3={\aG'},usedefault=@]{
  \ginfix{#1}{#2}{#3}{\gparop}
}
\newcommandx{\gcho}[3][1=i,2={\aG},3={\aG'},usedefault=@]{
  \ginfix{#1}{#2}{#3}{\gchoop}
}
\newcommandx{\gchov}[3][1=\gname,2={\aG},3={\aG'},usedefault=@]{
  \def\ggraph{\left(
  \begin{array}l
    \ifempty{#1}{{#2} \\ \gchoop \\ {#3}}{\!\!{#2} \\ \gchoop \\ {#3}}
  \end{array}\right)}
  \ifcp\gnode[{$#1$}][\ggraph] \else \ggraph \fi
}
\newcommandx{\grec}[3][1=i,2={\aG},3={\p},usedefault=@]{
  \def\ggraph{\grecop {#2} \ifempty{#3}{}{\grecopp {#3}}}
  \ifempty{#1}{\ggraph}{\gname[{$#1$}][\ggraph]}
}	

\newcommand{\getcentroid}[2]{
    \coordinate (tmpgatecoord) at (0,0);
    \foreach \n [count=\i] in {#1}{
      \path (\n);
      \coordinate (tmpgatecoord) at ($(tmpgatecoord) + (\n)$);
      \coordinate (#2) at ($1/\i*(tmpgatecoord)$);
}
}

\tikzset{
  gt/.style={
         ->,
         >=stealth',
         shorten >=.1pt,
         auto,
         node distance=3cm,
         scale = .5,
         every state/.style={inner sep = 2pt, minimum size = 0pt, font=\footnotesize, transform shape},
         transform shape
  },
lt/.style={
         ->,>=stealth',shorten >=1pt,auto,node distance=5cm,scale = 0.97,transform shape,
  },
  initial/.style={
         state,initial by arrow, initial text={}
  }
}

\tikzset{
  hgsem/.style={
    draw,
    node distance=2cm and 1cm,
    transform shape,
    smooth,
    every node/.style = {font=\sffamily\bfseries}
  }
}

\tikzset{
  cnode/.style={
    shape=circle,
    minimum size = 0mm,
    inner sep = 1pt,
    font=\tiny,
    draw
  },
  carrow/.style={
    ->,
    shorten >=1pt,
    >=stealth',
    auto,
    font=\scriptsize,
    draw,
    sloped
  }
}

\tikzset{
  CA/.style={
    transform shape,
	 node distance = 1.9cm,
    every state/.style = {cnode},
	 every edge/.style = {carrow}
  }
}

\tikzset{
  hgstyle/.style={
    src color={#1},
    tgt color={#1},
    centroid color={#1},
    centroid label={#1},
    centroid name={#1},
    centroid radius={#1},
    centroid ratio={#1},
    xoffset={#1},
    yoffset={#1},
    xsrcoffset={#1},
    ysrcoffset={#1},
    xtgtoffset={#1},
    ytgtoffset={#1},
    font={#1},
    centroid angle={#1},
    centroid tolerance={#1}
  },
src color/.store in = \hgsrccol,
  tgt color/.store in = \hgtgtcol,
  centroid color/.store in =\hgfillcolor,
  centroid label/.store in =\hglabel,
  centroid name/.store in =\hgname,
  centroid radius/.store in = \hgradius,
  centroid ratio/.store in = \hgratio,
  xoffset/.store in =\hgxoffset,
  yoffset/.store in =\hgyoffset,
  xsrcoffset/.store in =\hgxsrcoffset,
  ysrcoffset/.store in =\hgysrcoffset,
  xtgtoffset/.store in =\hgxtgtoffset,
  ytgtoffset/.store in =\hgytgtoffset,
  centroid angle/.store in =\hgangle,
  centroid tolerance/.store in =\hgtolerance,
src color = black,
  tgt color = black,
  centroid color = orange!40,
  centroid label={},
  centroid name={dummycentroid},
  centroid radius = .7pt,
  centroid ratio = .35,
  xoffset = 0,
  yoffset = 0,
  xsrcoffset = 0,
  ysrcoffset = 0,
  xtgtoffset = 0,
  ytgtoffset = 0,
  font=\sffamily\scriptsize,
  centroid angle=0,
  centroid tolerance=10pt
}

\newcommandx{\mkhg}[5][1={},4={},5={},usedefault=@]{
  \begingroup
  \tikzset{#1}
  \StrCount{#2,}{,}[\l] \StrCount{#3,}{,}[\m] \ifthenelse{\l = 1 \AND \m = 1}{
    \ifempty{#4}{
      \ifempty{#5}{
        \path[hgsem, ->, >=stealth', shorten >=1pt] (#2) -- (#3);
      }{
        \path[hgsem, ->, >=stealth', shorten >=1pt] (#2) #5 (#3);
      }
    }{
      \ifempty{#5}{
        \path[hgsem, ->, >=stealth', shorten >=1pt, #4] (#2) -- (#3);
      }{
        \path[hgsem, ->, >=stealth', shorten >=1pt, #4] (#2) #5 (#3);
      }
    }
  }{
    \coordinate (srcoffset) at (\hgxsrcoffset,\hgysrcoffset);
    \coordinate (tgtoffset) at (\hgxtgtoffset,\hgytgtoffset);
    \getcentroid{#2}{srccentroid};
    \getcentroid{#3}{tgtcentroid};
    \node[label={left:\hglabel}] (\hgname) at ($(srccentroid)!{1-\hgratio}!\hgangle:(tgtcentroid) + (\hgxoffset,\hgyoffset)$) {};
    \pgfgetlastxy \xc \yc;
    \pgfmathtruncatemacro{\xcontrol}{\xc};
    \pgfmathtruncatemacro{\ycontrol}{\yc};
    \foreach \n in {#2}{
      \path (\n);
      \pgfgetlastxy \xntmp \yntmp;
      \pgfmathtruncatemacro{\xn}{\xntmp};
      \pgfmathtruncatemacro{\yn}{\yntmp};
      \pgfmathsetmacro\xtmpdiff{abs(\xn - \xcontrol + \hgxsrcoffset)};
      \pgfmathsetmacro\ytmpdiff{abs(\yn - \ycontrol + \hgytgtoffset)};
      \ifdim \xtmpdiff pt > \hgtolerance
      \ifempty{#4}{
        \path[hgsem, \hgsrccol] (\n) .. controls ($(srccentroid.center) + (srcoffset)$) .. (\hgname.center);
      }{
        \path[hgsem, \hgsrccol] (\n) .. controls ($(srccentroid.center) + (srcoffset)$) .. (\hgname.center);
      }
      \else
      \ifempty{#4}{
        \path[hgsem, \hgsrccol] (\n) -- (\hgname.center);
      }{
        \path[hgsem, \hgsrccol, #4] (\n) -- (\hgname.center);
      }
      \fi
    }
    \foreach \n in {#3}{
      \path (\n);
      \pgfgetlastxy \xntmp \yntmp;
      \pgfmathtruncatemacro{\xn}{\xntmp};
      \pgfmathtruncatemacro{\yn}{\yntmp};
      \pgfmathsetmacro\xtmpdiff{abs(\xn - \xcontrol)};
      \pgfmathsetmacro\ytmpdiff{abs(\yn - \ycontrol)};
      \ifdim \xtmpdiff pt > \hgtolerance
      \ifempty{#4}{
        \path[hgsem, ->, >=stealth', shorten >=1pt, \hgtgtcol] (\hgname.center) .. controls (tgtcentroid.center) and ($(tgtcentroid.center) + (tgtoffset)$) .. (\n);
      }{
        \path[hgsem, ->, >=stealth', shorten >=1pt, \hgtgtcol,#4] (\hgname.center) .. controls (tgtcentroid.center) and ($(tgtcentroid.center) + (tgtoffset)$) .. (\n);
      }
      \else
      \ifempty{#4}{
        \path[hgsem, ->, >=stealth', shorten >=1pt, \hgtgtcol] (\hgname.center) --  (\n);
      }{
        \path[hgsem, ->, >=stealth', shorten >=1pt, \hgtgtcol] (\hgname.center) --  (\n);
      }
      \fi
    }
    \fill[\hgfillcolor] (\hgname) circle [radius=\hgradius];
  }
  \endgroup
}

\newcommandx{\hgordeq}[1][1={\aH},usedefault=@]{\sqsubseteq_{#1}}
\newcommandx{\gintsem}[4][4=.5]{
  \tikz[hgsem,scale=#4,every node/.style={font=\scriptsize}]{
    \node (out) {$\aout[{#1}][{#2}][][{#3}]$};
    \node[below = 20pt of out] (in) {$\ain[{#1}][{#2}][][{#3}]$};
    \mkhg{out}{in};
  }
}

\newcommandx{\gsem}[2][1={\aG},2={},usedefault=@]{\left\llbracket {#1} \right\rrbracket_{#2}}

\newcommandx{\rbot}{\text{undef}}
\newcommandx{\rtrs}[1][1={\aH},usedefault=@]{{#1}^{\star}}
\newcommandx{\gord}[1][1={\aG},usedefault=@]{\leq_{#1}}
\newcommandx{\gordeq}[1][1={\aG},usedefault=@]{\leq_{#1}}
\mkfun{\cause}{cs}{}
\mkfun{\effect}{ef}{}

\newcommandx{\aW}{w}
\newcommandx{\rlang}{\mathcal{L}}

\newcommand{\gfun}[1]{\ensuremath{\mathsf{\colorFun #1}}}
\mkfun{\eact}{\gfun{act}}{}
\mkfun{\enode}{\gfun{cp}}{}

\mkuop{\rmax}{\gfun{max}}{\aH}
\mkuop{\rmin}{\gfun{min}}{\aH}
\mkuop{\rMAX}{\gfun{lst}}{\aH}
\mkuop{\rMIN}{\gfun{fst}}{\aH}

\newcommandx{\rseq}[2][1=\aG,2={\aG'},usedefault=@]{\gfun{seq}({#1},{#2})}
\newcommandx{\rpar}[2][1=\aG,2={\aG'},usedefault=@]{\gfun{par}({#1},{#2})}

\newcommandx{\gproj}[2][1=\aG,2=\ptp]{{#1}\downarrow_{#2}}
\newcommandx{\cinit}[1][1={\aQzero},usedefault=@]{{#1}}
\newcommandx{\cfinal}[1][1={q_e},usedefault=@]{{#1}}

\newcommandx{\geproj}[4][1=\aG,2=\ptp,3=\cinit,4=\cfinal,usedefault=@]{
  {#1}\downarrow_{#2}^{{#3},{#4}}
}

\newcommand*{\StrikeThruDistance}{0.15cm}\tikzset{strike thru arrow/.style={
    decoration={markings, mark=at position 0.5 with {
        \draw [blue, thick,-] 
            ++ (-\StrikeThruDistance,-\StrikeThruDistance) 
            -- ( \StrikeThruDistance, \StrikeThruDistance);}
    },
    postaction={decorate},
}}

\newcommandx{\ich}[1][1={\aG},usedefault=@]{{#1}^{\oplus}}
\newcommandx{\ichedges}[2][1={\aG},2={\gname},usedefault=@]{{#1}^{\oplus}({#2})}
\newcommandx{\parts}[1]{2^{#1}}
\newcommandx{\actch}{c}
\newcommandx{\soundactch}[2][1={\aG},2={\actch},usedefault=@]{{#1} \,\circledR\, {#2}}
\newcommandx{\rOnActch}[2][1={\aG},2={\actch},usedefault=@]{{#1} \setminus {#2}}
\newcommandx{\rOnActchClean}[2][1={\aG},2={\actch},usedefault=@]{{#1} \circledR {#2}}
\newcommandx{\rAllEvents}[1][1={\aG},usedefault=@]{\mathit{dom}(#1)}

\newcommand{\AV}{\mathcal{V}}
\newcommand{\aH}{H}

\newcommandx{\hgvertex}[2][1=\al,2=\gname,usedefault=@]{{#1}_{\textcolor{red}{[{#2}]}}}
\newcommand{\aE}{{\colorE E}}
\renewcommand{\ae}[1][e]{{\colorE{#1}}}

\newcommand{\al}[1][l]{{\colorE{#1}}}
\newcommandx{\hyedge}[1]{\{#1\}}

\newcommandx{\rdiv}[2][1=\gcho,2=\ptp,usedefault=@]{
  \gfun{div}_{#2}(#1)
}

\newcommandx{\rrdiv}[5][1={\aG},2={\aG'},3={\AV},4={,\AV'},5=\ptp,usedefault=@]{
  \gfun{div}^{#3#4}_{#5}(#1,#2)
}
\newcommandx{\pdiv}[3][1={\apom_1},2={\apom_2},3={\apom},usedefault=@]{
  \gfun{div}_{#3}(#1,#2)
}
\newcommandx{\pfork}[3][1={\apom_1},2={\apom_2},3={\apom},usedefault=@]{
  \gfun{fork}_{#3}(#1,#2)
}

\tikzset{
  pomset/.style={
    node distance = .6cm and .6cm,
    scale = .7,
    transform shape,
    smooth
  }
}

\newcommandx{\mkint}[6][3=i,4=\p,5=\msg,6=\q,usedefault=@]{
\node[bblock, #1] (#2) {$\gint[#3][#4][#5][#6]$};
}

\newcommand{\mkseq}[2]{\path[line] (#1) -- (#2);}

\newcommand{\mknseq}[1]{
  \StrCount{#1}{,}[\l] \StrBefore{#1}{,}[\myhead]
  \StrBehind{#1}{,}[\mytail]
  \StrBefore{\mytail}{,}[\sndel]
  \ifnum \l > 1 {
    \mkseq{\myhead}{\sndel};
    \mknseq{\mytail}
  }
  \else{\ifnum \l > 0{
      \mkseq{\myhead}{\mytail};
    }
    \else{}
    \fi
  }
  \fi
}

\newcommandx{\mkgateblock}[6][6=yellow!10,usedefault=@]{
\path(#2);
  \pgfgetlastxy{\xgate}{\ygate};
  \pgfmathtruncatemacro{\xgateround}{\xgate};
  \StrCount{#3,}{,}[\l] \ifnum \l < 2 {\errmessage{#3 argument should be a comma-separated list of lenght >= 2}}
  \else{
    \foreach \n in {#3}{
      \path (\n);
      \pgfgetlastxy{\xnode}{\ynode};
      \pgfmathtruncatemacro{\xnround}{\xnode};
      \pgfmathsetmacro\tmpdiff{abs(\xnround - \xgateround)}
      \ifdim \tmpdiff pt > 1 pt \path[line] (#2) -| (\n);
      \else
        \path[line] (#2) -- (\n);
      \fi
    }
  }
  \fi
  \StrCount{#4,}{,}[\l] \ifnum \l < 2 {\errmessage{#4 argument should be a comma-separated list of lenght >= 2}}
  \else{
    \foreach \n in {#4}{
      \path (\n);
      \pgfgetlastxy{\xnode}{\ynode};
      \pgfmathtruncatemacro{\xnround}{\xnode};
      \pgfmathsetmacro\tmpdiff{abs(\xnround - \xgateround)}
      \ifdim \tmpdiff pt > 1 pt \path[line] (\n) |- (#5);
      \else
        \path[line] (\n) -- (#5);
      \fi
    }
  }
  \fi
  \node[#1] at (#2) {};
  \node[#1] at (#5) {};
  {
    \begin{pgfonlayer}{background}
      \path[fill=#6,rounded corners]
      (current bounding box.south west) rectangle
      (current bounding box.north east);
    \end{pgfonlayer}
  }
}

\newcommandx{\mkbranchblock}[5][5=@]{
  \mkgateblock{ogate}{#1}{#2}{#3}{#4}[#5]
}

\newcommandx{\mkforkblock}[5][5=@]{
  \mkgateblock{agate}{#1}{#2}{#3}{#4}[#5]
}

\newcommandx{\mkgraph}[3][1=.5cm, usedefault=@]{
  \node[source,above = #1 of {#2}] (src#2) {};
  \node[sink,below  = #1 of {#3}] (sink#3) {};
  \path[line] (src#2) -- (#2);
  \path[line] (#3) -- (sink#3);
}

\newcommandx{\mkloop}[5][1=.5, 2=1.5, 5=\aguard, usedefault=@]{
\node[lgate,above = #1 of {#3}] (entry#3) {};
  \pgfgetlastxy \xentry \yentry;
  \pgfmathtruncatemacro{\xentryrounded}{\xentry};
  \node[below = #1 of {#4}, label = {above right:{$#5$}},yshift=-1em] (dummy) {};
  \node[lgate,below  = #1 of {#4}] (exit#4) {};
  \pgfgetlastxy \xexit \yexit;
  \pgfmathtruncatemacro{\xexitrounded}{\xexit};
  \path[line] (entry#3) -- (#3);
  \path[line] (#4) -- (exit#4);
  \pgfmathsetmacro\tmpdiff{abs(\xentryrounded - \xexitrounded)}
  \path[line, color=teal] (exit#4) -| ($(exit#4)+(\tmpdiff,0)+(#2,0)$) |- (entry#3);
}

\newcommandx{\mkfork}[5][2=gatenode,3=i,4=.6,5=2,usedefault=@]{
  \mkgatebegin{#1}[{\gname[{#3}]}][agate][#4]{#2}[#5]
}

\newcommandx{\mkbranch}[5][2=gatenode, 3=i, 4=.6, 5=2, usedefault=@]{
  \mkgatebegin{#1}[{\gname[{#3}]}][ogate][#4]{#2}[#5]
}

\newcommandx{\mkgatebegin}[6][2={}, 3=ogate, 4=.5, 6=2, usedefault=@]{
\coordinate (gatecoord) at (0,0);
  \coordinate (xmax) at (0,0);
  \coordinate (xmin) at (0,0);
  \pgfgetlastxy \xmin \xmax;
  \foreach \n [count=\i] in {#1}{
    \pgfgetlastxy \xc \yc;
    \path (\n);
    \pgfgetlastxy \xn \yn;
    \ifnum \i = 1
      \coordinate (xmin) at (\xn,0);
      \coordinate (xmax) at (\xn,0);
      \coordinate (max) at (0,\yn);
    \else
      \ifdim \xn < \xmin
        \coordinate (xmin) at (\xn,0);
      \fi
      \ifdim \xn > \xmax
        \coordinate (xmax) at (\xn,0);
      \fi
      \ifdim \yn < \yc
        \coordinate (max) at (0,\yc);
      \else
        \coordinate (max) at (0,\yn);
      \fi
    \fi
  }
  \coordinate (gatecoord) at ($(xmin)!.5!(xmax) + (max) + (0,#4) + (max)$);
  \node[#3,label={below:$#2$}] (#5) at (gatecoord) {};
  \pgfgetlastxy{\xgate}{\ygate};
  \pgfmathtruncatemacro{\xgateround}{\xgate};
  \StrCount{#1,}{,}[\l] \ifnum \l < 2 {\errmessage{#1 argument should be a comma-separated list of lenght >= 2}}
  \else{
    \foreach \n in {#1}{
      \path (\n);
      \pgfgetlastxy{\xnode}{\ynode};
      \pgfmathtruncatemacro{\xnround}{\xnode};
      \pgfmathsetmacro\tmpdiff{abs(\xnround - \xgateround)}
      \ifdim \tmpdiff pt > #6 pt {\path[line] (#5) -| (\n);}
      \else
        \path[line] (#5) -- (\n);
      \fi
    }
  }
  \fi
}

\newcommandx{\mkgatebeginold}[5][2={},3=ogate,4=.5,usedefault=@]{
\coordinate (gatecoord) at (0,0);
  \foreach \n [count=\i] in {#1}{
    \pgfgetlastxy \xc \yc;
    \path (\n);
    \pgfgetlastxy \xn \yn;
    \coordinate (gatecoord) at ($(gatecoord) + (\xn,0)$);
    \coordinate (gatecoord) at ($1/\i*(gatecoord)$);
    \ifdim \yn < \yc
    \node (max) at (0,\yc) {};
    \else
    \node (max) at (0,\yn) {};
    \fi
  }
  \coordinate (gatecoord) at ($(gatecoord) + (0,#4) + (max)$);
  \node[#3,label={below:$#2$}] (#5) at (gatecoord) {};
  \pgfgetlastxy{\xgate}{\ygate};
  \pgfmathtruncatemacro{\xgateround}{\xgate};
  \StrCount{#1,}{,}[\l] \ifnum \l < 2 {\errmessage{#1 argument should be a comma-separated list of lenght >= 2}}
  \else{
    \foreach \n in {#1}{
      \path (\n);
      \pgfgetlastxy{\xnode}{\ynode};
      \pgfmathtruncatemacro{\xnround}{\xnode};
      \pgfmathsetmacro\tmpdiff{abs(\xnround - \xgateround)}
      \ifdim \tmpdiff pt > 1 pt \path[line] (#5) -| (\n);
      \else
        \path[line] (#5) -- (\n);
      \fi
    }
  }
  \fi
}

\newcommandx{\mkmerge}[5][2=gatenode,3=i,4=.5,5=2,usedefault=@]{
  \mkgateend{#1}[{\ifempty{#3}{}{\nmerge[#3]}}][ogate][#4]{#2}[#5]
}

\newcommandx{\mkjoin}[5][2=gatenode,3=i,4=.5,5=2,usedefault=@]{
  \mkgateend{#1}[{\ifempty{#3}{}{\nmerge[#3]}}][agate][#4]{#2}[#5]
}

\newcommandx{\mkgateend}[6][2={},3=ogate,4=.5,6=2,usedefault=@]{
\coordinate (gatecoord) at (0,0);
  \coordinate (xmax) at (0,0);
  \coordinate (xmin) at (0,0);
  \pgfgetlastxy \xmin \xmax;
  \foreach \n [count=\i] in {#1}{
    \pgfgetlastxy \xc \yc;
    \path (\n);
    \pgfgetlastxy \xn \yn;
    \ifnum \i = 1
      \coordinate (xmin) at (\xn,0);
      \coordinate (xmax) at (\xn,0);
      \coordinate (ymin) at (0,\yn);
    \else
      \ifdim \xn < \xmin
        \coordinate (xmin) at (\xn,0);
      \fi
      \ifdim \xn > \xmax
        \coordinate (xmax) at (\xn,0);
      \fi
      \ifdim \yn > \yc
        \coordinate (ymin) at (0,\yc);
      \else
        \coordinate (ymin) at (0,\yn);
      \fi
    \fi
  }
  \coordinate (gatecoord) at ($(xmin)!.5!(xmax) + (ymin)$);
  \node[#3,label={above:$#2$}] (#5) at ($(gatecoord) - (0,{#4})$) {};
  \pgfgetlastxy{\xgate}{\ygate};
  \pgfmathtruncatemacro{\xgateround}{\xgate};
  \StrCount{#1,}{,}[\l] \ifnum \l < 2 {\errmessage{#1 argument should be a comma-separated list of lenght >= 2}}
  \else{
    \foreach \n in {#1}{
      \path (\n);
      \pgfgetlastxy{\xnode}{\ynode};
      \pgfmathtruncatemacro{\xnround}{\xnode};
      \pgfmathsetmacro\tmpdiff{abs(\xnround - \xgateround)}
      \ifdim \tmpdiff pt > #6 pt {\path[line] (\n) |- (#5);}
      \else
        \path[line] (\n) -- (#5);
      \fi
    }
  }
  \fi
}

\newcommandx{\mkgateendold}[5][2={},3=ogate,4=.5,usedefault=@]{
\coordinate (gatecoord) at (0,0);
  \coordinate (xmax) at (0,0);
  \coordinate (xmin) at (0,0);
  \pgfgetlastxy \xmin \xmax;
  \foreach \n [count=\i] in {#1}{
    \pgfgetlastxy \xc \yc;
    \path (\n);
    \pgfgetlastxy \xn \yn;
    \ifdim \xn < \xmin
    \coordinate (xmin) at (\xn,0);
    \fi
    \ifdim \xn > \xmax
    \coordinate (xmax) at (\xn,0);
    \fi
    \ifdim \yn > \yc
    \coordinate (ymin) at (0,\yc);
    \else
    \coordinate (ymin) at (0,\yn);
    \fi
    \coordinate (gatecoord) at ($(xmin)!.5!(xmax) + (ymin)$);
  }
  \node[#3,label={above:$#2$}] (#5) at ($(gatecoord) - (0,{#4})$) {};
  \pgfgetlastxy{\xgate}{\ygate};
  \pgfmathtruncatemacro{\xgateround}{\xgate};
  \StrCount{#1,}{,}[\l] \ifnum \l < 2 {\errmessage{#1 argument should be a comma-separated list of lenght >= 2}}
  \else{
    \foreach \n in {#1}{
      \path (\n);
      \pgfgetlastxy{\xnode}{\ynode};
      \pgfmathtruncatemacro{\xnround}{\xnode};
      \pgfmathsetmacro\tmpdiff{abs(\xnround - \xgateround)}
      \ifdim \tmpdiff pt > 1 pt \path[line] (\n) |- (#5);
      \else
        \path[line] (\n) - (#5);
      \fi
    }
  }
  \fi
}

\newcommand{\gatedistancein}{3pt}
\newcommand{\gatedistanceinand}{2pt}

\usetikzlibrary{
  arrows,
  backgrounds,
  chains,
  calc,
  decorations.markings,decorations.pathreplacing,
  fadings,
  fit,
  patterns,
  petri,
  positioning,
  shadows,
  shapes,automata,shapes.callouts
}

\tikzset{
  src/.style={draw,circle,fill=white,
    minimum size=2mm,
    inner sep=0pt
  },
  sink/.style={draw,circle,double,fill=white,
    minimum size=1.5mm,
    inner sep=0pt
  },
  node/.style={draw,circle,fill=black,
    minimum size=2mm,
    inner sep=0pt
  },
source/.style={draw,circle,fill=white,
    minimum size=3mm,
    inner sep=0pt
  },
  sink/.style={draw,circle,double,fill=white,
    minimum size=3mm,
    inner sep=0pt
  },
block/.style = {rectangle, draw=gray, align=center, fill=orange!25, rounded corners=0.1cm,
    minimum size=5mm, inner sep=2pt},
  prenode/.style = {minimum size=9pt,inner sep=2pt, font=\Large},
bblock/.style = {rectangle, draw=blue!50, opacity=.7, line width=.5pt, align=center, fill=white, rounded corners=0.1cm,
    minimum size=4mm, inner sep=1pt},
  prenode/.style = {minimum size=9pt,inner sep=2pt, font=\Large},
agate/.style={draw, rectangle,
    minimum size=3mm,
    inner sep=0pt,
    fill=orange!25,
    label={[red]center:$\mid$}
  },
ogate/.style = {
    diamond, draw, fill=orange!25,
    minimum size=4mm,
    inner sep=0pt,
    label={[red]center:$+$}
  },
lgate/.style = {
    diamond, draw, fill=orange!25,
    minimum size=4mm,
    inner sep=0pt,
    label={[red]center:$\circlearrowleft$}
    },
altogate/.style = {
    diamond, draw,
    minimum size=4mm,
    inner sep=0pt,
    postaction={path picture={\draw
        ([yshift=\gatedistancein]path picture bounding box.south) -- ([yshift=-\gatedistancein]path picture bounding box.north)
        ([xshift=-\gatedistancein]path picture bounding box.east) -- ([xshift=\gatedistancein]path picture bounding box.west)
        ;}}},
  altgate/.style={draw, rectangle,
    minimum size=3mm,
    inner sep=0pt,
    postaction={path picture={\draw
        ([yshift=\gatedistanceinand]path picture bounding box.south) --
        ([yshift=-\gatedistanceinand]path picture bounding box.north) ;}}},
anygate/.style = {circle, draw, fill=white,
    minimum size=4mm,
    inner sep=0pt,
    postaction={path picture={\draw[black]
        ([xshift=-\gatedistancein,yshift=\gatedistancein]path picture bounding box.south east) --
        ([xshift=\gatedistancein,yshift=-\gatedistancein]path picture bounding box.north west)
        ([xshift=-\gatedistancein,yshift=-\gatedistancein]path picture bounding box.north east) --
        ([xshift=\gatedistancein,yshift=\gatedistancein]path picture bounding box.south west)
        ;}}
  },
smallglobal/.style={
        node distance=1cm and 0.8cm, semithick, scale=0.8, every node/.style={transform shape}
  },
elli/.style = {draw,densely dotted,-},
line/.style = {draw,->, rounded corners=0.07cm,>=latex},
  nline/.style = {draw,semithick, ->},
  pline/.style = {draw,->,>=latex},
  node distance=1cm and 0.7cm,
  baseline=(current  bounding  box.center),
  local/.style={rectangle, draw, fill=\fillcolor, drop shadow,
    text centered, rounded corners, minimum height=5em
  },
  bigar/.style={
    draw,very thick, ->
  },
  process/.style={rectangle, fill=\fillcolor, drop shadow, align=center,
    text centered, text=gray},
  choreo/.style={rectangle, fill=\fillcolor, drop shadow, align=center,
    text centered, rounded corners},
  machinecloud/.style={
    cloud, cloud puffs=10, cloud ignores aspect, minimum height=.1cm, minimum width=2cm, draw
  },
  fitting node/.style={
    inner sep=0pt,
    fill=none,
    draw=none,
    reset transform,
    fit={(\pgf@pathminx,\pgf@pathminy) (\pgf@pathmaxx,\pgf@pathmaxy)}
  },
  mypetri/.style={
    font=\footnotesize,
    baseline=(current  bounding  box.center)
  },
  silentrans/.style = {rectangle, draw=black, align=center, fill=black,
    minimum height=1pt,
    minimum width=15pt,
    inner sep=1.5pt
  },
  reset transform/.code={\pgftransformreset},
  tmtape/.style={draw,minimum size=1.2cm}
}

\newcommand{\gunlessop}{\mbox{\colorOp\tiny\tt unless}}

\newcommandx{\gtry}[5][1=\gname,2={\aG_1 \gchoop \cdots \gchoop \aG_n},3=\gin,4=\gout,5={j},usedefault=@]{
  \def\foo{\gtryop\ {#2} \ \gcatchop\ {#3} {\colorOp \Rightarrow} {#4} {\colorOp \bullet} {\gname[{#5}]}}
  \gnode[{#1}][{\ifempty{#1} {\foo } {(\foo)}}]
}

\newcommandx{\gtrycatch}[4][1=\gname,2={\aG},3=\gin,4={\aG'},usedefault=@]{
  \def\foo{\gtryop\ {#2} \ \gcatchop\ {#3} \gdoop\ {#4}}
  \gnode[{#1}][{\ifempty{#1} {\foo} {(\foo)}}]
}

\newcommandx{\agG}[2][1={\aG},2=\aguard]{{#1} \ifempty{#2}{}{\ \gunlessop\ {#2}}}

\newcommandx{\grcho}[5][1=\gname,2={\agG},3={\agG[\aG'][\aguard']},4={\cdots},5=A,usedefault=@]{
  \def\foo{{#2} {\ \ifempty{#4}{\gchoop}{\gchoop \ifempty{#4}{}{\ {#4}\  \gchoop}}\ } {#3}}
  \ifempty{#1}{\ifempty{#5}{\foo}{\gselop\ \cpt[{#1}][{\ptp[#5]}]\big\{ \foo \big\}}}{\gselop\ \cpt[{#1}][{\ptp[#5]}]\big\{ \foo \big\}}
}

\newcommandx{\ggprefix}[3][1=\ptp,2={\aR},3={\aR'},usedefault=@]{f_{#1}} \newcommand{\aconfigfn}{\chi}
\newcommand{\aconfig}{\ell}

\newcommand{\lstates}{\statemap}
\newcommandx{\sysconfig}[3][1=\lstates,2=\aconfigfn,3={},usedefault=@]{
  \conf{ {#1},{#2} \ifempty{#3}{}{, #3} }
}
\newcommand{\sysctxfn}[1][]{\gamma_{#1}}
\newcommandx{\sysctx}[2][1=\aQ,2={},usedefault=@]{({#1},\sysctxfn[{#2}])}

\newcommandx{\alog}[4][1=\msg,2=q,3=\gname,4=t,usedefault=@]{({#1},{#2},{#3},{#4})}

\newcommand{\aCM}{\ensuremath{M}}\newcommand{\aM}{\aCM}
\newcommand{\aQ}{Q}
\newcommandx{\aQzero}[1][1=,usedefault=@]{
  {\ifempty{#1}{q_0}{q_{0#1}}}
}
\newcommand{\badbranches}[1][]{\beta\ifempty{#1}{}{({#1})}}
\newcommand{\aTrs}{\tset}
\newcommandx{\guardedaction}[2][1=\al,2=\aguard,usedefault=@]{
  {#1} \ifempty{#2}{}{/} {#2}
}
\newcommandx{\atrM}[4][1=q,2=\al,3={\hat q,\hat \al, \aguard},4=q',usedefault=@]{
  {#1} \xrightarrow[{#3}]{\guardedaction[{#2}][]} {{#4}}
}
\newcommandx{\atrS}[5][
  1={\sysconfig[@][@][\badbranches]},
  2=\al,
  3=\aguard,
  4={\sysconfig[\lstates'][\aconfigfn'][\badbranches]},
  5=\sysctx,usedefault=@
]{
  {#1} \xRightarrow{\qquad} {{#4}}
}
\newcommandx{\arevtrS}[2][
  1={\sysconfig[@][@][\badbranches]},
  2={\sysconfig[\lstates'][\aconfigfn'][\badbranches']},
  usedefault=@
]{
  {#1} \rightsquigarrow {#2}
}
\newcommand{\aCS}{\ensuremath{S}}

\newcommandx{\enables}[2][1=\aconfigfn,2=\aguard,usedefault=@]{{#1} \vdash {#2}}
\newcommandx{\gprojfn}[5][1=\aG,2=\ptp,3=\cinit,4=\cfinal,5={},usedefault=@]{
  \mathbf{proj}_{#2}({#1},{#3},{#4}\ifempty{#5}{}{,{#5}})
}

\newcommandx{\rbp}[3][1=\aG,2=\aconfigfn,3=\achan,usedefault=@]{\mathtt{RBP}_{{#1},{#2}}\ifempty{#3}{}{({#3})}}

\newcommand{\apseudoCFSM}{\mathtt{M}}
\newcommandx{\pseudoseq}[2][1=\apseudoCFSM,2=\apseudoCFSM',usedefault=@]{{#1}  ; {#2}}
\newcommandx{\pseudoCFSM}[4][1=\aQ,2=\aQzero,3=\cfinal,4=\aTrs,usedefault=@]{(#1 \ ; #2 \ ; #3 \ ; #4)}
\newcommandx{\markt}[3][1=\hat{\al},2=\hat{q},3=\aguard,usedefault=@]{\%\big({#1} , {#2}, {#3}\big)}

\newcommandx{\borderfn}[2][1=\aconfig,2=\aloop,usedefault=@]{
  \mathsf{border}_{{#2}}\ifempty{#1}{}{({#1})}
}

\tikzset{
  mycallout/.style={
	 fill=gray!30, opacity=.5, overlay, align=center,
	 cloud callout, cloud puffs=10, aspect=1.9, cloud ignores aspect, cloud puff arc=100
  }
}

\newcommandx{\ggvisually}[8][1=5pt,2=15pt,3=5pt,4=5pt,5=1.0cm,6=\scriptsize,7={},8={},usedefault=@]{
\def\dist{\hspace{#5}}
  $\begin{array}{c@{\dist}c@{\dist}c@{\dist}c@{\dist}c@{\dist}c}
\begin{tikzpicture}[node distance=0.9cm and 0.4cm, every node/.style={scale=.7,transform shape}]
		 \node[source] (srcint) {};
		 \node[sink,below=of srcint] (sinkint) {};
		 \node[mycallout, above = .3cm of srcint, xshift=1cm, callout absolute pointer={(srcint.east)}] {source node};
		 \node[mycallout, below = .3cm of sinkint, xshift=1cm, callout absolute pointer={(sinkint.west)}] {sink node};
		 \path[line] (srcint) -- (sinkint);
	  \end{tikzpicture}
	  &
\begin{tikzpicture}[node distance=0.9cm and 0.4cm, every node/.style={scale=.7,transform shape}]
			\mkint{}{int}[]
			\mkgraph{int}{int};
\end{tikzpicture}
	  &
\begin{tikzpicture}[node distance=.9cm and 0.4cm, every node/.style={scale=.7,transform shape}]
			\node[bblock] at (0,0) (g) {$\aG$};
			\node[node, below=of g] (s1) {};
			\node[bblock, below=of s1] (gp) {$\aG'$};
			\path[line,dotted] (g) -- (s1);
			\path[line,dotted] (s1) -- (gp);
		 \end{tikzpicture}
	  &
\begin{tikzpicture}[node distance=.4cm and 0.4cm, every node/.style={scale=.7,transform shape}]
			\node[bblock] at (-.7,0) (g) {$\aG$};
			\node[bblock] at (.7,0)  (gp) {$\aG'$};
			\node[node, above=of g] (f) {};
			\node[node, below=of g] (j) {};
			\node[node, above=of gp] (fp) {};
			\node[node, below=of gp] (jp) {};
			\path[line,dotted] (f) -- (g);
			\path[line,dotted] (g) -- (j);
			\path[line,dotted] (fp) -- (gp);
			\path[line,dotted] (gp) -- (jp);
			\mkfork{f,fp}[fork][][#1];
			\mkjoin{j,jp}[join][][#2];
			\mkgraph{fork}{join};
			\node[mycallout, above = .3cm of fork, xshift=-1cm, callout absolute pointer={(fork.west)}] {fork gate};
			\node[mycallout, above = -.9cm of join, xshift=-1cm, callout absolute pointer={(join.west)}] {join gate};
		 \end{tikzpicture}
	  &
\begin{tikzpicture}[node distance=.4cm and 0.4cm, every node/.style={scale=.7,transform shape}]
			\node[bblock] at (-.7,0) (g) {$\aG$};
			\node[bblock] at (.7,0)  (gp) {$\aG'$};
			\node[node, above=of g] (f) {};
			\node[node, below=of g] (j) {};
			\node[node, above=of gp] (fp) {};
			\node[node, below=of gp] (jp) {};
			\path[line,dotted] (f) -- (g);
			\path[line,dotted] (g) -- (j);
			\path[line,dotted] (fp) -- (gp);
			\path[line,dotted] (gp) -- (jp);
			\mkbranch{f,fp}[fork][][#3];
			\mkmerge{j,jp}[join][][#4];
         \mkgraph{fork}{join};
         \node[mycallout, above = .3cm of fork, xshift=-1cm, callout absolute pointer={(fork.west)}] {branch gate};
         \node[mycallout, above = -.9cm of join, xshift=-1cm, callout absolute pointer={(join.west)}] {merge gate};
       \end{tikzpicture}
     \ifempty{#7}{}{
     &
\begin{tikzpicture}[node distance=0.4cm and 0.4cm, every node/.style={scale=.7,transform shape}]
        \node[bblock] (g) {$\aG$};
        \node[node, above=.5cm of g] (f) {};
        \node[node, below=.5cm of g] (j) {};
        \path[line,dotted] (f) -- (g);
        \path[line,dotted] (g) -- (j);
        \mkloop[.4][1]{f}{j};
        \mkgraph[.3cm]{entryf}{exitj};
        \node[mycallout, above = .2cm of entryf, xshift=1.3cm, callout absolute pointer={(entryf.east)}] {loop entry};
        \node[mycallout, above = -.7cm of exitj, xshift=1.3cm, callout absolute pointer={(exitj.west)}] {loop exit};
      \end{tikzpicture}
     }
     \ifempty{#8}{}{
	  \\
     \text{#6 empty}
     &
     \text{#6 interaction}
     &
     \text{#6 sequential}
     &
     \text{#6 parallel}
     &
     \text{#6 branch}
&\text{#6 iteration}
   }
   \end{array}$
}

\newcommandx{\newggvisually}[5][1=5pt,2=15pt,3=\scriptsize,4={},5={},usedefault=@]{
\def\w{1cm}
  \begin{minipage}[c]{\w}
	 \ifempty{#5}{}{\text{#3 empty}\\[#1]}
	 \begin{tikzpicture}[node distance=0.9cm and 0.4cm, every node/.style={scale=.7,transform shape}]
		\node[source] (srcint) {};
		\node[sink,below=of srcint] (sinkint) {};
		\node[mycallout, above = .3cm of srcint, xshift=1cm, callout absolute pointer={(srcint.east)}] {source node};
		\node[mycallout, below = .3cm of sinkint, xshift=1cm, callout absolute pointer={(sinkint.west)}] {sink node};
		\path[line] (srcint) -- (sinkint);
	 \end{tikzpicture}
  \end{minipage}
  \hfill
\begin{minipage}[c]{\w}
     \ifempty{#5}{}{\text{#3 interaction}\\[#1]}
		 \begin{tikzpicture}[node distance=0.9cm and 0.4cm, every node/.style={scale=.7,transform shape}]
			\mkint{}{int}[]
			\mkgraph{int}{int};
\end{tikzpicture}
	  \end{minipage}
	 \hfill
\begin{minipage}[c]{.1cm}
     \ifempty{#5}{}{\text{#3 sequential}\\[#1]}
		 \begin{tikzpicture}[node distance=.9cm and 0.4cm, every node/.style={scale=.7,transform shape}]
			\node[bblock] at (0,0) (g) {$\aG$};
			\node[node, below=of g] (s1) {};
			\node[bblock, below=of s1] (gp) {$\aG'$};
			\path[line,dotted] (g) -- (s1);
			\path[line,dotted] (s1) -- (gp);
		 \end{tikzpicture}
	  \end{minipage}
	  \hfill
\begin{minipage}[c]{\w}
     \ifempty{#5}{}{\text{#3 parallel}\\[#1]}
		 \begin{tikzpicture}[node distance=.4cm and 0.4cm, every node/.style={scale=.7,transform shape}]
			\node[bblock] at (-.7,0) (g) {$\aG$};
			\node[bblock] at (.7,0)  (gp) {$\aG'$};
			\node[node, above=of g] (f) {};
			\node[node, below=of g] (j) {};
			\node[node, above=of gp] (fp) {};
			\node[node, below=of gp] (jp) {};
			\path[line,dotted] (f) -- (g);
			\path[line,dotted] (g) -- (j);
			\path[line,dotted] (fp) -- (gp);
			\path[line,dotted] (gp) -- (jp);
			\mkfork{f,fp}[fork][][#1];
			\mkjoin{j,jp}[join][][#2];
			\mkgraph{fork}{join};
			\node[mycallout, above = .2cm of fork, xshift=-1cm, callout absolute pointer={(fork.west)}] {#3 fork gate};
			\node[mycallout, below = .2cm of join, xshift=-1cm, callout absolute pointer={(join.west)}] {#3 join gate};
		 \end{tikzpicture}
	  \end{minipage}
	  \hfill
\begin{minipage}[c]{\w}
     \ifempty{#5}{}{\text{#3 branch}\\[#1]}
		 \begin{tikzpicture}[node distance=.4cm and 0.4cm, every node/.style={scale=.7,transform shape}]
			\node[bblock] at (-.7,0) (g) {$\aG$};
			\node[bblock] at (.7,0)  (gp) {$\aG'$};
			\node[node, above=of g] (f) {};
			\node[node, below=of g] (j) {};
			\node[node, above=of gp] (fp) {};
			\node[node, below=of gp] (jp) {};
			\path[line,dotted] (f) -- (g);
			\path[line,dotted] (g) -- (j);
			\path[line,dotted] (fp) -- (gp);
			\path[line,dotted] (gp) -- (jp);
			\mkbranch{f,fp}[fork][][#1];
			\mkmerge{j,jp}[join][][#2];
         \mkgraph{fork}{join};
         \node[mycallout, above = .1cm of fork, xshift=-1cm, callout absolute pointer={(fork.west)}] {branch gate};
         \node[mycallout, below = .1cm of join, xshift=-1cm, callout absolute pointer={(join.west)}] {merge gate};
       \end{tikzpicture}
	  \end{minipage}
     \ifempty{#4}{}{
     \hfill
\begin{minipage}[c]{\w}
     \ifempty{#5}{}{\text{#3 iteration}\\[#1]}
		 \begin{tikzpicture}[node distance=0.4cm and 0.4cm, every node/.style={scale=.7,transform shape}]
        \node[bblock] (g) {$\aG$};
        \node[node, above=.5cm of g] (f) {};
        \node[node, below=.5cm of g] (j) {};
        \path[line,dotted] (f) -- (g);
        \path[line,dotted] (g) -- (j);
        \mkloop[.4][1]{f}{j};
        \mkgraph[.3cm]{entryf}{exitj};
        \node[mycallout, above = .2cm of entryf, xshift=1.3cm, callout absolute pointer={(entryf.east)}] {loop entry};
        \node[mycallout, above = -.7cm of exitj, xshift=1.3cm, callout absolute pointer={(exitj.east)}] {loop exit};
      \end{tikzpicture}
	  \end{minipage}
     }
}

  \newcommandx{\wwwcquote}[1][1=quo:w3c,usedefault=@]{
	 \ifempty{#1}{}{\begin{quote}\label{#1}}
		\lq\lq Using the Web Services Choreography specification, a
		\textcolor{orange}{contract} containing a global definition of the
		common \textcolor{orange}{ordering conditions and constraints}
		under which \textcolor{orange}{messages} are exchanged, is
		produced that describes, from a \textcolor{orange}{global
		  viewpoint} [...]  observable behaviour of all the parties
		involved.
\textcolor{OliveGreen}{Each party} can then use the global definition to
		\textcolor{OliveGreen}{build and test solutions that conform to it}.
The global specification is in turn \textcolor{OliveGreen}{realised by combination of} the
		resulting \textcolor{orange}{local systems} [...]\rq\rq
		\ifempty{#1}{}{\end{quote}}
  }

 \newcommand{\thetool}{\toolid{TRAC}} \newcommand{\states}{\mathsf{S}}
\newcommand{\finalStates}{\mathsf{F}}
\newcommand{\stateQ}[1][]{\mathsf{q}_{#1}}

\newcommand{\partyP}[1][]{\mathsf{p}_{#1}}
\newcommand{\partyPi}[1][]{\mathsf{p'}_{#1}}
\newcommand{\partyB}{\mathsf{b}}
\newcommand{\function}[1]{\mathsf{#1}}
\newcommand{\functionF}[1][]{\function{f}_{#1}}
\newcommand{\functionFi}[1][]{\function{f'}_{#1}}
\newcommand{\functionC}[1]{\contractC.\function{#1}}
\newcommand{\functionCG}{\contractC.\function{g}}
\newcommand{\functionCF}[1][]{\functionC{\functionF[#1]}}
\newcommand{\functionCFi}[1][]{\functionC{\functionFi[#1]}}
\newcommand{\contractC}[1][]{\mathsf{c}_{#1}}
\newcommand{\contractCi}[1][]{\mathsf{c'}_{#1}}
\newcommand{\var}[1]{\mathsf{#1}}
\newcommand{\varX}[1][]{\var{x}_{#1}}
\newcommand{\varY}[1][]{\var{y}_{#1}}
\newcommand{\varC}[1]{\contractC.\var{#1}}
\newcommand{\varCX}[1][]{\varC{\varX[#1]}}
\newcommand{\varCY}[1][]{\varC{\varY[#1]}}
\newcommand{\guardG}[1][]{\mathsf{g}_{#1}}
\newcommand{\guardGi}[1][]{\mathsf{g'}_{#1}}
\newcommand{\exprE}[1][]{\mathsf{e}_{#1}}
\newcommand{\roleR}[1][]{\mathsf{R}_{#1}}
\newcommand{\roleRi}[1][]{\mathsf{R'}_{#1}}
\newcommand{\old}{\mathsf{old}}
\newcommandx{\binder}[3][1=\mathsf{any}, 2=\partyP, 3=\roleR, usedefault=@]{{#1}\ifempty{#2}{}{\;{#2} \colon {#3}}}
\newcommand{\any}{\binder}
\newcommand{\new}{\binder[\nu]}
\newcommand{\modParties}{\mathcal{P}}
\newcommand{\Guards}{\mathcal{G}}
\newcommand{\modparty}[1][]{\pi_{#1}}

\newcommand{\Vars}[1][]{\mathcal{V}_{#1}}
\newcommand{\Contracts}{\mathcal{C}}
\newcommand{\init}[1][]{\mathsf{start}}
\newcommand{\Functions}[1][]{\mathcal{F}_{#1}}
\newcommand{\assign}[1][]{\beta_{#1}}
\newcommand{\Assignments}{\mathcal{B}}
\newcommand{\Assign}[1][]{B_{#1}}
\newcommand{\Assigni}[1][]{B'_{#1}}
\newcommand{\fin}{\mathsf{fin}}
\newcommand{\decl}[1][]{\mathsf{d}_{#1}}
\newcommand{\Decl}{\mathcal{D}}

\newcommand{\stateS}[1][]{\mathsf{s}_{#1}}
\newcommand{\stateSi}[1][]{\mathsf{s'}_{#1}}
\newcommand{\modelname}{DAFSM\xspace}
\newcommand{\modelnames}{DAFSMs\xspace}
\newcommand{\roleO}{\mathsf{O}}
\newcommand{\roleB}{\mathsf{B}}
\newcommand{\partyO}{\mathsf{o}}
\newcommand{\pathS}[1][]{\sigma_{#1}}
\newcommand{\dafsm}[1][]{\mathcal{S}_{#1}}
\newcommand{\transT}[1][]{\mathsf{t}_{#1}}

\newcommand{\subst}[2]{\{#1/#2\}}
\newcommand{\dataVars}{\mathsf{dataParams}}
\newcommand{\IntType}{\mathsf{Int}}
\newcommand{\partyConfl}{\#}
\newcommand{\labelset}{\mathcal{L}}
\newcommand{\alabel}[1][]{\ell_{#1}}
\newcommand{\txt}[1]{\texttt{#1}}
\tikzset{
  dafsm/.style={
	 ->, >=stealth', auto, semithick,
	 every edge/.style={draw,sloped},
	 every state/.style={
		fill=white,
		draw=black,
		thick,
		text=black,
		scale=1,
		minimum size = 0mm,
		inner sep = 2pt
	 }
  }
}

\newcommand{\toUpdate}[1]{\textcolor{Bittersweet}{#1}}
\renewcommand{\toUpdate}[1]{{#1}}

\newcommand{\startingFile}{\texttt{TXT}}

\newcommand{\validator}{\texttt{Validator}\xspace}

\newcommand{\trGrinder}{\texttt{TrGrinder}\xspace}
\newcommand{\graphGen}{\texttt{GraphGen}\xspace}
\newcommand{\callerChecker}{\texttt{CallerCheck}\xspace}
\newcommand{\detChecker}{\texttt{\toUpdate{DetCheck}}\xspace}
\newcommand{\aConsistency}{\texttt{AConsistency}\xspace}
\newcommand{\formulaBuilder}{\texttt{FBuilder}\xspace}
\newcommand{\ztModel}{\texttt{Z3Model}\xspace}
\newcommand{\ztRunner}{\texttt{Z3Runner}\xspace}
\newcommand{\ztModelAnalyzer}{\texttt{Analizer}\xspace}
\newcommand{\analysisVerdict}{\texttt{Verdict}\xspace}
\newcommand{\fsmImage}{\texttt{V-FSM}\xspace}

\def\mkNode#1{\begin{tikzpicture}
		\useasboundingbox (-1ex,0)rectangle(1ex,0);
		\node[anchor=base,draw,fill=black,text=white,shape=circle,font=\footnotesize\bfseries,inner sep=1pt]{#1};
\end{tikzpicture}}

\newcommand{\feature}[1]{\textsf{#1}}
\newcommand{\ok}{\textcolor{OliveGreen}{\checkmark}}
\newcommand{\ko}{\textcolor{BrickRed}{\textsf{x}}}
\newcommand{\na}{$\ominus$}
\newcommand{\ib}{\rotatebox[origin=c]{180}{\ok}}

 \usepackage[firstpage]{draftwatermark}

\SetWatermarkText{\hspace*{-2.2in}\raisebox{-9in}{\href{https://eapls.org/pages/artifact_badges/}{\mbox{\includegraphics[width=1.4cm]{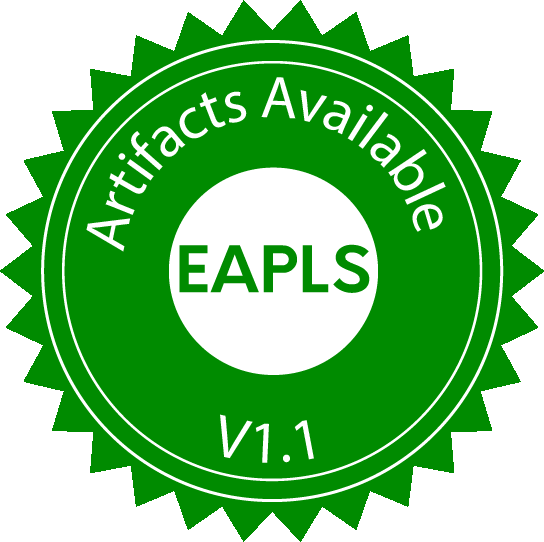}\includegraphics[width=1.4cm]{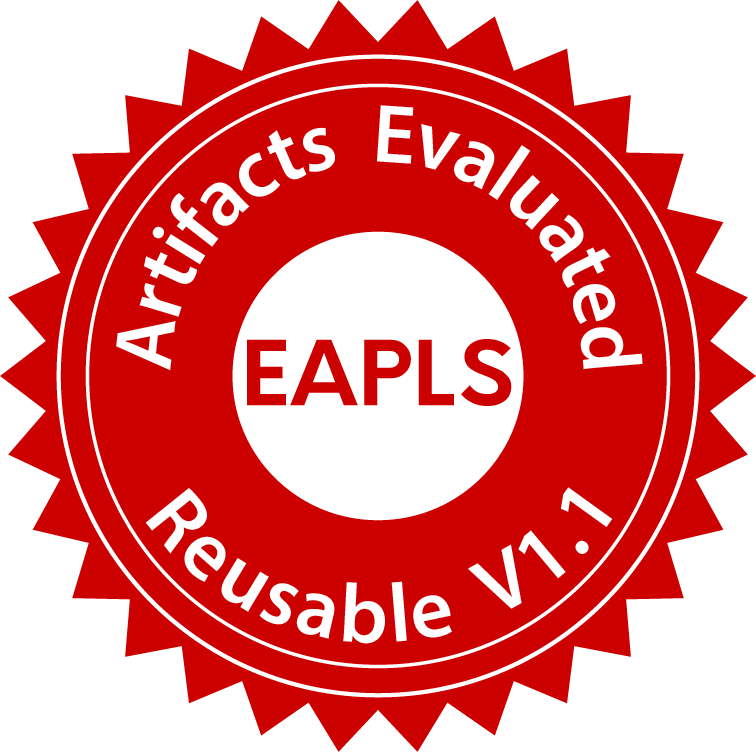}}}}}
\SetWatermarkAngle{0}

\title{
\thetool: a tool for data-aware coordination
}
\subtitle{(with an application to smart contracts)}
\author{
  Jo\~ao Afonso\inst{1}\and
  Elvis Konjoh Selabi\inst{2,3}\and
  Maurizio Murgia\inst{3}\and\\
  Ant\'onio Ravara\inst{1}\and
  Emilio Tuosto\inst{3}}\institute{
  NOVA School of Science and Technology
  \and
  Universit\`a di Camerino
  \and
  Gran Sasso Science Institute
}

\makeatletter
\begin{document}

\maketitle

\begin{abstract}
  We propose \thetool, a tool for the specification and verification
  of coordinated multiparty distributed systems.
Relying on finite-state machines (FSMs) where transition labels look
  like Hoare triples, \thetool can specify the coordination of the
  participants of a distributed protocol for instance an execution
  model akin blockchain smart contracts (SCs).
In fact, the transitions of our FSMs yield guards, and assignments
  over data variables, and with participants binders.
The latter allow us to model scenarios with an unbounded number of
  participants which can vary at run-time.
We introduce a notion of \emph{well-formedness} to rule out
  meaningless or problematic specifications.
This notion is verified with \thetool and demonstrated on several
  case studies borrowed from the smart contracts domain.
Then, we evaluate the performance of \thetool using a set of
  randomised examples, studying the correlations between the features
  supported and the time taken to decide well-formedness.
\end{abstract}

\section{Introduction}\label{sec:intro}
\newcommand{\tracgithub}{\url{https://github.com/loctet/TRAC}}
We propose \thetool, a tool to support the coordination
of distributed applications.
The design of \thetool is inspired by the Azure initiative of
Microsoft~\cite{azure} which advocates the use of
finite-state machines (FMSs) to specify the
coordination of smart contract (SC for short).
This idea is not formalised; in fact, Azure's FSMs are informal
sketches aiming to capture the \quo{correct} executions of SCs.
For instance, the FSM for the simple market place (SMP) scenario
borrowed from~\cite{azure/smp} (the textual description is ours):
\\
\begin{tikzpicture}
 \useasboundingbox (3cm,2cm)rectangle(-3.3cm,-1.8cm);
  \node (smp) {\includegraphics[width=6cm]{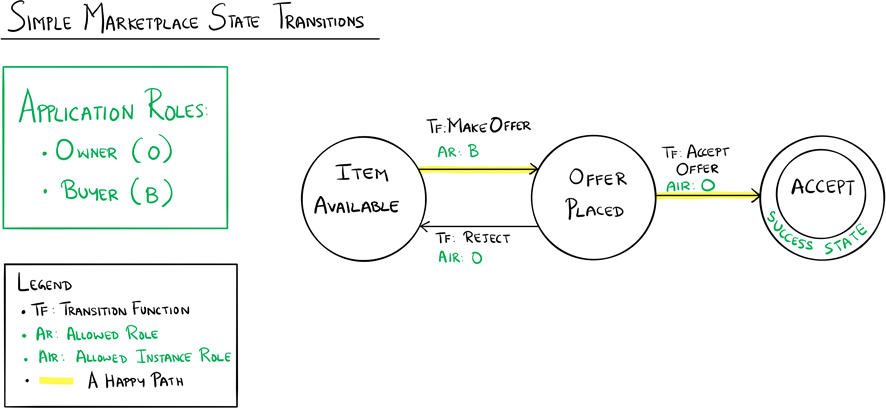}};
  \node (txt) [right = .5cm of smp, fill=orange!20, opacity=.5, yshift=.2cm, text width = 6.5cm, scale = .8, align=left]{The sketch declares the roles (Owner and Buyer) played by participants.
	 
	 In the initial state \texttt{Item Available} the buyer is allowed
	 to make an offer, moving the protocol to the \texttt{Offer Placed}
	 state where two options are possible: the owner either accepts the
	 offer (making the protocol reache the success state
	 \texttt{Accept}) or rejects the offer (moving back the protocol to
	 \texttt{Item Available}).

	 The labels of the transitions specify which role executes with
	 operations to make the protocol progress.
  };
\end{tikzpicture}
\\
The FSM informally specifies a protocol coordinating the participants
enacting the roles owner and buyer, from a \emph{global} standpoint;
we call \emph{coordination protocol} such specification.
A coordination protocol can be regarded as \emph{global view} --in the
sense of choreographies~\cite{bpmn,w3c:cho}-- where the state of the
protocol determines which operations are enabled.
This resembles the execution model of monitors~\cite{han73}.
In fact, as in monitors, coordination protocols encapsulate a state
that --through an API-- concurrent processes can have exclusive access
to.
The API is basically a set of operations guarded by conditions set to
maintain an invariant on the encapsulated state (in the SMP scenario
the operations are \texttt{MakeOffer}, \texttt{AcceptOffer}, and
\texttt{Reject}).
The key differences between coordination protocols and
monitors~\cite{han73} is that in the former ($i$) participants are
distributed and do not share memory, ($ii$) the invocation of an
operation whose guards is not valid in the current state is simply
ignored without preempting the caller, and therefore ($iii$) processes
do not have to be awaken.

We aim to refine the approach of Azure so to enable algorithmic
verification of relevant properties of data-aware coordination of
protocols.
In fact, as for monitors, the interplay among the operations that
modify the state and the guards in the API can lead to unexpected
behaviours when informal specifications are used.
We illustrate this problem with some examples on the SMP example.
\begin{enumerate}
\item The sketch of SMP does not clarify if a participant can play
  more roles simultaneously; for instance, it is not clear if an owner
  must be a different instance than buyers.
\item\label{it:newold} The labels distinguish roles and instances
  (\texttt{AR} and \texttt{AIR}): in fact, it is assumed that there
  can be many instances of a same role.
Scope and quantification of roles is not clear; for instance, a
  requirement specified in~\cite{azure/smp} reads \quo{The transitions
	 between the \texttt{Item Available} and the \texttt{Offer Placed}
		states can continue until the owner is satisfied with the offer
		made.}
This sentence does not clarify if, after a rejection, the new offer
  can be made by a new buyer or it must be the original one;
\item The sketch specify neither the conditions enabling operations in
  a given state nor how operations change the state of the contract's
  variables; should the price of the item remain unchanged when the
  owner invokes the \texttt{Reject}?
\end{enumerate}

\noindent\textbf{Contributions \& Structure}
\cref{sec:model} introduces \emph{data-aware FSMs} (\modelnames) to
formalise coordination protocols.
Roughly, \modelnames allow specifications ($i$) to express conditions
on how operations affect the state of the protocol and ($ii$) to
explicitly declare the capabilities of participants.
We propose \emph{well-formedness} condition on \modelnames to rule out
erroneous coordination protocols.

The definition of \modelnames is instrumental to our main contribution
which is \thetool, a tool realising our model described in
\cref{sec:thetool}.
We build on an initial proposal developed in~\cite{afonso23}.

The applicability of \thetool is demonstrated by showing how its
features can specify and verify the SCs in~\cite{azure}.
Moreover, we discuss the performances of the \thetool with an
experimental evaluation (cf. \cref{sec:expl}).
\MMcomm{The source code of \thetool and our experimental data is available at \tracgithub.}

Related work and conclusions are given respectively in \cref{sec:rw,sec:conc}.

\section{Data-aware FSMs}\label{sec:model}

In our model, protocols' \emph{participants} cooperate through a
\emph{coordinator} according to their \emph{role}.
We let $\partyP,\partyPi,\hdots$ denote \emph{participant variables},
$\roleR,\roleRi,\hdots$ denote \emph{roles}, and $\contractC,\contractCi,\hdots$
denote $\Contracts$ \emph{coordinator names}.
Each coordinator name $\contractC$ has:
\begin{itemize}
\item A finite set $\Vars[\contractC]$ of \emph{data variables}; we
  let $\varCX,\varCY,\hdots$ range over $\Vars[\contractC]$ and write
  $\varX,\varY,\hdots$ when the coordinator name is clear
  from the context.
Each variables has an associated data type, \eg Int, Bool,
  $\hdots$; we also allow usual structured data types like arrays.
\item A set of \emph{function names}, ranged over by
  $\functionCF,\functionCFi,\hdots$.
Function parameters, ranged over by $\varX,\varY,\hdots$, can be
  either data or participants variables; we allow function calls with
  different parameters to a same function.
\end{itemize}
An \emph{assignment} takes the form $\varCX := \exprE$, where $\exprE$
is an expression;\footnote{We borrow from Z3 a substantial subset of expressions over variables
  and parameters (barred participants parameters) whose syntax is
  standard and therefore omitted.
We assume that expressions do not have side effects.
} the set $\Assignments$ of assignments is ranged over by $\assign$
while $\Assign,\Assigni,\hdots$ range over finite subsets of $\assign$
where each variable can be assigned at most once; moreover, we assume
that all assignments in $\Assign$ are executed simultaneously.
In an assignment $\varCX := \exprE$ data variables occurring in
$\exprE$ must have the $\old$ qualifier to refer to the value of
$\varCX$ before the assignment.
The set of \emph{guards} $\Guards$, ranged over by
$\guardG,\guardGi, \hdots$, consists of constraints (i.e., boolean
expressions) over data variables and function parameters.
Parameter \emph{declarations} are written as $\varX:T$ or
$\partyP:\roleR$ to respectively assign data type $T$ to $\varX$ and
role $\roleR$ to $\partyP$; we let $\Decl$ be the set of all
declarations and $\decl$ to range over $\Decl$.
Lists of declarations are denoted by $\vec\decl$ with the implicit
assumption that the parameters in $\vec\decl$ are pairwise distinct.

The set $\modParties$ of \emph{qualified participants} consists of the
terms generated by
\[
\modparty \bnfdef \new \bnfmid \any \bnfmid \partyP
\]
where both $\new[]$ and $\any[@][]$ are binders.
Intuitively, $\new$ specifies that variable $\partyP$ represents a
fresh participant with role $\roleR$ while $\any$ qualifies $\partyP$
as an existing participant with role $\roleR$.
With $\partyP$ we refer to a participant in the scope of a binder.

Before its formal definition (cf.~\cref{def:dafsm}), we give
an intuitive account of our model.
We use FSMs as coordination protocols with a single coordinator~$\contractC$.
The transitions of an FSM represent the call to functions exposed by
the coordinator $\contractC$ performed by participants.
Such calls may update the current control state (by means of state
transitions)
{ and the state of data variables (by mean of assignments).}
Access to functions can be restricted to some participants (using
participants variables and modifiers), and the availability of a
function may depend on the current control or data states (using
guards).
A protocol starts in the initial state of the FSM specifying where the
initial state of variables is set by the creator of the coordinator;
intuitively, the creator may be thought of as an object in
object-oriented programming created by invoking a constructor.

\begin{definition}\label{def:dafsm}
  Let $2^{\Assignments}_{\fin}$ be the set of all finite subsets of
  $\Assignments$ and
  $\labelset = \Guards \times \modParties \times \Functions \times
  \vec{\Decl} \times 2^{\Assignments}_{\fin}$ be the set of
  \emph{labels}, ranged over by $\alabel$.
A \emph{data-aware finite state machine} (\modelname for short) is a
  tuple
  $\dafsm = (\states,\stateQ[0],\trans{},\finalStates,
  \contractC,\new,\vec{\decl[0]},\Assign[0])$ where:
\begin{itemize}
\item $(\states, \stateQ[0], \trans{}, \finalStates)$ is an FSM over
  $\labelset$ (namely, $\states$ is finite set of \emph{states},
  $\stateQ[0] \in \states$ is the \emph{initial state},
  $\trans{} \subseteq \states \times \labelset \times \states$, and
  $\finalStates \subseteq \states$ is the set of \emph{accepting}
  states);
\item $\contractC \in \Contracts$ is the coordinator name;
\item for each transition label
  $(\guardG, \modparty, \functionF, \vec\decl, \Assign)$, if
  $\varCX := \exprE \in \Assign$ then every data parameter occurring in $\exprE$ occurs in
  $\vec{\decl}$, $\exprE$ is well typed, and the data
  variables occurring in the guards of any of the transitions of
  $\dafsm$ belong to $\Vars[\contractC]$;
\item $\new$ binds $\partyP$ to the participant creating the
  coordinator;
\item $\vec{\decl[0]} \subseteq \vec{\Decl}$ is the parameters list of
  the coordinator;
\item $\Assign[0] \subseteq_{\fin} \Assignments$ is a set of
  assignments (setting the initial values of the state variables).
\end{itemize}
A \emph{path} is a finite sequence of transitions
$\stateS[0] \trans{\alabel[1]} \stateS[1] \cdots \stateS[n]
\trans{\alabel[n]} \stateS[n+1]$ with $\stateS[0] = \stateQ[0]$.
\end{definition}

The next example introduces a convenient graphical notation for
\modelnames in which guards on transitions are in curly brackets for
readability; this notation is reminiscent of Hoare triples (guards are
not to be confused with sets).
\begin{example}\label{ex:smp}
Let 
$\alabel[\text{new}] = \{\var{offer} > 0\}\ \new[\partyB][\roleB]
  \triangleright \functionC{makeOffer}(\IntType:\var{offer})\
  \{\varC{offer} := \var{offer}\}$ and
  $\alabel[\text{ext}] = \{\var{offer} > 0\}\ \any[@][\partyB][\roleB]
  \triangleright \functionC{makeOffer}(\IntType:\var{offer})\
  \{\varC{offer} := \var{offer}\}$.
The \modelnames below represents the SMP protocol of \cref{sec:intro}.

  \begin{center}
  \begin{tikzpicture}[dafsm, node distance = 4cm]
		\node (dummy) {};
		\node[state, right = of dummy] (q0) {$\stateQ[0]$};
		\node[state] (q1) [below of=q0, yshift=2.7cm] {$\stateQ[1]$};
		\node[state] (q1') [left of=q1] {$\stateQ[1]'$};
		\node[state,accepting] (q2) [right of=q1]  {$\stateQ[2]$};
		\path (dummy) edge[left] node[above, text width=3.9cm, align=center] {
		  $\new[\partyO][\roleO]\triangleright\init(\contractC,\IntType:\var{price})$
		  $\{\varC{price} := \var{price}\}$} (q0);
		\path (q0) edge node[rotate=90, xshift=.3cm, yshift=-.2cm] {$\alabel[\text{new}]$} (q1);
		\path (q1) edge node {$\partyO \triangleright \functionC{acceptOffer}()$} (q2);
		\path (q1) edge[right] node[sloped, anchor=center, above] {$\partyO \triangleright \functionC{rejectOffer}()$}
		(q1')
		(q1') edge[bend right=20] node[sloped, anchor=center, below, near start]{$\alabel[\text{ext}]$}
		(q1.south west)
		(q1') edge[bend left=30] node[sloped, anchor=center, above, near start] {$\alabel[\text{new}]$} (q1.north west);
	 \end{tikzpicture}
  \end{center}
The initial state is $\stateQ[0]$ and it is graphically represented
  by the source-less arrow entering it.
The label\footnote{We may omit writing guards when they are $\mathsf{True}$ and
	 assignments when they are empty as in the transitions from $\stateQ[1]$.
} of this arrow represents the invocation from a new participant
  $\partyO$ with the owner's role $\roleO$ to the constructor for a
  coordinator $\contractC$ with a parameter $\var{price}$ of type
  $\IntType$.
The set of assignments is the singleton initialising the coordinator's variable
  $\varC{price}$ to $\var{price}$.
  
  In $\stateQ[0]$, the only enabled function is
  $\functionC{makeOffer}(\IntType:\var{offer})$; the first buyer
  $\partyB$ invoking this function with a parameter $\var{offer}$
  satisfying the guard $\var{offer} > 0$ moves the protocol to state
  $\stateQ[1]$ while recording the new offer in the coordinator state
  with the assignment $\varC{offer} := \var{offer}$.
Contextually, the state of the coordinator records that the caller
  $\partyB$ plays role $\roleB$.

  From state $\stateQ[1]$ only the owner $\partyO$ can make the protocol progress
  by either accepting or rejecting the offer.
In the former case, the protocol reaches the accepting state
  $\stateQ[2]$ (graphically denoted with a doubly-circled node); in
  the latter case, the protocol reaches state $\stateQ[1]'$ where
  either an existing buyer or a new one can make further offers.
\finex
\end{example}
Notably, the DAFSM of \cref{ex:smp} refines the informal one in
\cref{sec:intro} by more precisely specifying that offers can arrive
either from previous buyers or new ones (cf. item~\ref{it:newold} in
\cref{sec:intro}).

\subsection{Well-formedness of \modelname}
The restrictions in \cref{def:dafsm} concern single transitions;
however, \modelnames can model meaningless and wrong behaviours, due
to conditions spanning several transitions, \eg free occurrences of
participant variables, lack of participants of a role or inconsistent
guards. Below we spell-out those constraints after motivating them
with simple examples.

\medskip
A first issue is the presence of free occurrences of participants names.
\begin{example}\label{ex:free-party}
  The \modelname \raisebox{6pt}{\tikz[node distance= 4cm, dafsm, scale=.85, transform shape]{
		\node[state] (S0)      {$\stateS[0]$};
		\node[state,draw=none] (S_) [left of=S0] {};
		\node[state,accepting] (S1) [right of=S0] {$\stateS[1]$};
		
		\path
		(S_) edge node {
			$\new[\partyO][\roleO]\triangleright\init(\contractC)$} (S0)
		(S0) edge node {
			$\partyP\triangleright\functionCF()$}
		 (S1);
}
}  is syntactically erroneous since the participant variable $\partyP$
	is not bound.
\finex
\end{example}

In our model qualified participants of the form $\new$ and $\any$, and
parameter declarations of the form $\partyP:\roleR$ act as binders.
In a \modelname all occurrences of participant variable should be in
the scope of a binder to be meaningful.
Formally, we say that a transition
$(\stateS[1],\guardG,\modparty,\functionCF,\vec{\decl},
\assign,\stateS[2])$ binds $\partyP$ iff:
\[
\exists \roleR: \modparty = \new[\partyP][\roleR]\quad \lor\quad \modparty = \any \quad
\lor\quad \partyP:\roleR \in \vec{\decl}
\]
The occurrence of $\partyP$ in a path
$\pathS = \pathS[1]
(\stateS[1],\guardG,\partyP,\functionCF,\vec{\decl},\assign,\stateS[2])
\pathS[2]$ is \emph{bound in $\pathS$} if there is a transition in
$\pathS[1]$ binding $\partyP$ and it is \emph{bound in a \modelname
  $\dafsm$} if all the paths of $\dafsm$ including the occurrence binding
it.
Finally, $\dafsm$ is \emph{closed} if all occurrences of participant
variables are bound in $\dafsm$.

\medskip
Another problem arises when the role of a qualified participant is empty.
\begin{example}
  If we bind the occurrence of $\partyP$ in the \modelname of
  \Cref{ex:free-party} with the binder $\any[@][]$, we obtain the
  closed \modelname
  \[\dafsm[2] = \raisebox{6pt}{\begin{tikzpicture}[node distance= 4cm, dafsm]
		
		\node[state] (S0)      {$\stateS[0]$};
		\node[state,draw=none] (S_) [left of=S0] {};
		\node[state,accepting] (S1) [right of=S0] {$\stateS[1]$};
		
		\path
		(S_) edge node {
			$\new[\partyO][\roleO]\triangleright\init(\contractC)$} (S0)
		(S0) edge node {
			$\any\triangleright\functionCF()$}
		 (S1);
	  \end{tikzpicture}}
	\]
However, we argue that $\dafsm[2]$ is ill-formed since $\roleR$ is
	necessarily empty in $\stateS[0]$.
Hence no action is possible, and the execution gets stuck in the
	initial state.
\finex
\end{example}

We now propose a simple syntactical check that avoids the problem of empty roles. Notice 
that a sound and complete procedure for empty roles detection subsumes reachability, which
may be undecidable depending on the chosen expressivity of constraints and expressions.

A binder \emph{expands} role $\roleR$ if it is a qualified participant
of the form $\new[\partyP][\roleR]$ or a parameter declaration of the
form $\partyP:\roleR$.
A role $\roleR$ is \emph{expanded} in a path $\pathS$ iff:
\[
\pathS = \pathS[1] (\stateS[1],\guardG,\any[@][\partyP][\roleR],\functionCF,\vec{\decl},
\assign,\stateS[2]) \pathS[2]
\implies \exists \transT \in \pathS[1]: \transT\;\text{expands}\; \roleR
\]
A \modelname $\dafsm$ \emph{expands} a role $\roleR$ if
every path of $\dafsm$ expands $\roleR$.
Finally, $\dafsm$ is (\emph{strongly}) \emph{empty-role free} if
$\dafsm$ expands every role in $\dafsm$.

Despite the quantification over the possibly infinite set of all paths, empty-role freedom
can be decided by considering only \emph{acyclic} paths, that is paths which contain at most one 
occurrence of each state. Clearly, there are only finitely many acyclic paths.
Notice that $\dafsm[2]$ above is not empty-role free.

\medskip

Finally, progress can be jeopardised if assignments falsify
all the guards of the subsequent transitions.
\begin{example}
  The \modelname $\dafsm[3]$ below is both closed and empty-role free,
  as the caller of $\functionCF$ is $\partyO$ which is bound by the
  constructor, and there are no $\any[@][]$ modifiers.
  \[\toUpdate{ \dafsm[3] = }\raisebox{6pt}{\begin{tikzpicture}[node distance= 4.5cm, dafsm]
		
		\node[state] (S0)      {$\stateS[0]$};
		\node[state,draw=none] (S_) [left of=S0] {};
		\node[state,accepting] (S1) [right of=S0] {$\stateS[1]$};
		
		\path
		(S_) edge node {
			$\new[\partyO][\roleO]\triangleright\init(\contractC)\;
			\{\varCX := 0\}$} (S0)
		(S0) edge node {
			$\{\mathsf{\varCX > 0}\}\;\partyO\triangleright\functionCF()$}
		 (S1);
	  \end{tikzpicture}
	}
	\]
	Crucially, $\varCX > 0$ will never be satisfied at run-time because
	$\varCX$ is initialised to $0$ and never changed.
So again every execution gets stuck in state $\stateS[0]$.
\finex
\end{example}

Similarly to empty roles, detecting inconsistencies is undecidable at
least for expressive enough constraints and expressions. We therefore
devise a syntactic technique amenable of algorithmic verification. The
idea is to check that every transition $\transT$, regardless of the
\quo{history} of the current execution, leads to a state which is either
accepting or it has at least a transition enabled. This is intuitively
accomplished by checking that the guard of $\transT$, after being
updated according to the assignments of $\transT$, implies the
disjunction of the guards of the outgoing transitions from the target
state of $\transT$.  Before formally introduce our notion of
consistency, we need a few auxiliary definitions.

\begin{definition}
  For all states $\stateS$, we define the \emph{progress constraint}
  $\guardG[\stateS]$ as $\mathsf{True}$ when $\stateS$ is accepting,
  and as the disjunction of guards of the outgoing transitions of
  $\stateS$.
Let $\varCX \not\in \Assign$ mean
  that for all $\varCY := \exprE \in \Assign$, $\varCY$ and $\varCX$
  differ and the expression $\exprE$ is not $\varCX$.
The \emph{progress constraint} of an assignment $\Assign$ is
  \[
	 \guardG[\Assign] = \bigwedge_{(\varCX := \exprE) \in \Assign}
	 \varCX = \exprE \land \bigwedge_{\varCX \not\in \Assign} \varCX =
	 \old\; \varCX
  \]
  We define $\dataVars(\vec{\decl})$ as the list of data parameter
  names occurring in $\vec{\decl}$.
\end{definition}

We can now define our notion of consistency.
\begin{definition}\label{def:consistency}
  Let $\guardG\subst{\vec \varY}{\vec \varX}$ be the guard obtained
  from $\guardG$ after the simultaneous substitution of variables
  $\vec \varX$ with $\vec \varY$. A transition
  $\transT =
  (\stateS,\guardG,\modparty,\functionCF,\vec{\decl},\assign,\stateSi)$
  is \emph{consistent} if:
\[
\forall \vec {\varCX}, \vec{\old\; \varCX}: \exists \dataVars(\vec{\decl}):
(\guardG\subst{\vec{\old\; \varCX}}{\vec {\varCX}} \land \guardG[\Assign]) \implies \guardG[\stateSi] 
\]
A \modelname $\dafsm$ is \emph{consistent} if so is every
transition of $\dafsm$.
\end{definition}

\begin{example}\label{ex:usingold}
	The \modelname $\dafsm[4]$ below shows the importance of renaming variable with $\old$. The \aConsistency formula of $\dafsm[4]$ for the transition from $\stateS[0]$ to $\stateS[1]$ is $\forall \varCX : True \quad \& \quad \varCX = \varCX + 1 => True $. The latter formula is evaluated as $ False \implies True $ which is $True$. We don't want this inconsistency case, therefore, by replacement, the \aConsistency formula of $\dafsm[4]$ becomes  $\forall \varCX, \varCX_\old : True \quad \& \quad \varCX = \varCX_\old + 1 => True $.
	\[\toUpdate{ \dafsm[4] = }\raisebox{6pt}{\begin{tikzpicture}[node distance= 3cm, dafsm]
			
			\node[state] (S0)      {$\stateS[0]$};
			\node[state,draw=none] (S_) [left of=S0] {};
			\node[state] (S1) [right=4.3cm of S0] {$\stateS[1]$};
			\node[state,accepting] (S2) [right of=S1] {$\stateS[2]$};
			
			\path
			(S_) edge node {
				$\new[\partyO][\roleO]\triangleright\init(\contractC)\;$} (S0)
			(S0) edge node {
				$\{\mathsf{True}\}\;\partyO\triangleright\functionCF[1]() \{\varCX := \varCX + 1\}$} (S1)
			(S1) edge node[distance= 4cm] {
				$\{\mathsf{True}\}\;\partyO\triangleright\functionCF[2]()$}
			(S2);
		\end{tikzpicture}
	}
	\]
\finex
\end{example}
Non-determinism could be useful for some applications, most of the
time \emph{determinism} is a desirable property (e.g., SCs are usually
required to be deterministic~\cite{ethereum_white_paper}).
Before the formal definition, we give a few examples illustrating how
non-determinism may arise in \modelnames.

\begin{example}
  The \modelname
  $\dafsm = \raisebox{10pt}{
		\begin{tikzpicture}[dafsm, node distance= .5cm and 2cm]
		\node[state] (S0)      {$\stateS[0]$};
		\node[state,draw=none] (dummy) [above =of S0] {};
		\node[state] (S1) [left =of S0] {$\stateS[1]$};
		\node[state,accepting] (S2) [right =of S0] {$\stateS[2]$};
\path
		(dummy) edge node[rotate = 90,left,yshift=.2cm] {
			$\new[\partyO][\roleO]\triangleright\init(\contractC)$} (S0)
		(S0) edge[bend right=10] node {$\alabel[1]$} (S1)
		(S1) edge[bend right=10] node[sloped, anchor=center, below]{$\partyO\triangleright\functionCG()$} (S0)
		(S0) edge node[above] {$\alabel[2]$} (S2);
    \end{tikzpicture}
	 }
  $
  is deterministic or not, depending on the labels $\alabel[1]$
  and $\alabel[2]$.
Let us consider some cases.
  \begin{description}
  \item[{
		$\alabel[1] = \alabel[2] = \partyO\triangleright\functionCG()$}]
	 $\dafsm$ is non-deterministic because a call to function
	 $\functionCF$ by $\partyO$ can lead either to $\stateS[1]$ or to
	 $\stateS[2]$.
  \item[{
		$\alabel[1] = \new[\partyP][\roleR]\triangleright\functionCG()$
		and
		$\alabel[2] =
		\any[@][\partyP][\roleR]\triangleright\functionCG()$}] $\dafsm$
	 is deterministic intuitively because the next state is
	 unambiguously determined by the caller of $\functionCG$: the protocol
	 moves to $\stateS[1]$ or $\stateS[2]$ depending whether the call
	 is performed by an existing or a new participant.
  \item[{
	 $\alabel[1] = \{\varX \leq
	 10\}\;\partyO\triangleright\functionCG(\varX:\IntType)$ and
	 $\alabel[2] = \{\varX >
	 10\}\;\partyO\triangleright\functionCG(\varX:\IntType)$}]
	 $\dafsm$ is deterministic because guard $\varX \leq 10$ leading to
	 $\stateS[1]$ and guard $\varX > 10$ leading to $\stateS[2]$ are
	 disjoint; therefore the next state is determined by the value of
	 the parameter $\varX$, and every value enables at most one
	 transition.
  \end{description}
  Also, taking $\alabel[1]$ as in the latter case and
  $\alabel[2] = \{\varX \geq
  10\}\;\partyO\triangleright\functionCG(\varX:\IntType)$ would make
  $\dafsm$ non-deterministic because the guards of $\alabel[1]$ and of
  $\alabel[2]$ are not disjoint therefore the next state is not
  determined by the caller of $\functionCG$.
\finex
\end{example}

We now define a notion of \emph{strong determinism}, which is
decidable and can be efficiently established.
To this aim, we first define the binary relation
$\partyConfl \subseteq \modParties \times \modParties$ as the least
symmetric relation satisfying:
\[
  \new[{\partyP}][{\roleR}]\; \partyConfl\; \partyP',
  \quad
  \new[{\partyP}][{\roleR}]\; \partyConfl\; \any[@][\partyP'][\roleO],
  \qand
  \roleR \neq \roleO \implies \any[@][\partyP][\roleR]\; \partyConfl\; \any[@][\partyP'][\roleO]
\]
Intuitively, if $\modparty[1]\;\partyConfl\;\modparty[2]$, then the
callers in $\modparty[1]$ and $\modparty[2]$ \emph{differ}.
Indeed, the first two item just say that a new participant is
necessarily different from an existing one. The third item says that
two participant with different roles are necessarily different (since
we require that every participant can have at most one role).

We now define strong determinism.
\begin{definition}\label{def:determinism}
  A \modelname $\dafsm$ is \emph{(strongly) deterministic} if, for all
  transitions $\transT[1] \neq \transT[2]$ in $\dafsm$ such that
  $\transT[1]$ and $\transT[2]$ have the same source state \toUpdate{and the same function}  then:
  \[
	 (\guardG[1] \land \guardG[2] \implies \mathsf{False}) \quad \lor
	 \quad \modparty[1] \partyConfl \modparty[2]
  \]
  where, for $i \in \{1,2\}$, $\guardG[i]$ is the guard of $\transT[i]$
  and $\modparty[i]$ is the qualified participant of $\transT[i]$.
\end{definition}

A \modelname is \emph{well-formed} when empty-role free,
consistent, and deterministic.

\section{The Tool}\label{sec:thetool}
We implement our model in \thetool.
Specifically, \thetool renders \modelnames in terms of a DSL to
specify \modelnames and verify the well-formedness condition defined
in~\cref{sec:model} relying on the SMT solver Z3.
We present the architecture of \thetool in \cref{sec:arch} and some
implementation details in \cref{sec:impl}.

\subsection{Architecture}\label{sec:arch}
\tikzset{
	every node/.style={
		rectangle,
		align=center,
		thin,
		scale=.8,
		font=\ttfamily
	},
	every path/.style={
		draw,
		thick,
		rounded corners,
		-latex
	},
	used/.style={
		top color=white,
		anchor=center,
		bottom color=red!50!black!20
	},
	data/.style={
		top color = orange,
		bottom color = yellow!50!black!20,
		sharp corners
	},
	output/.style={
		top color = black,
		rounded corners,
		bottom color = black!90!black!10,
		trapezium, trapezium left angle=-120, trapezium right angle=-60,
		color = white
	},
	dev/.style={
		draw,
		semithick,
		rounded corners,
		anchor=center,
		drop shadow={color=blue!15, shadow scale = .95},
		top color=white,
		bottom color=blue!50!black!20,
		font=\ttfamily
	},
	num/.style={
		anchor=center,
		fill=black,
		text=white,
		circle,
inner sep=0pt,
		font=\bfseries
	}
}
\begin{figure}[t]\centering  \setlength{\belowcaptionskip}{-1cm}
	\begin{tikzpicture}[node distance = .6cm and 1.1cm]
\node(txt) [data, minimum width=1.3cm] {\startingFile};
		\node(ggen) [used, right = of txt] {\graphGen};
		\node(vald) [dev, below = of txt, minimum width=1.3cm, yshift = -0.2cm] {\validator};
		\node(call) [dev, below right = of ggen] {\callerChecker};
		\node(z3) [data, right = of call, yshift=.8cm, xshift = 0.17cm] {\ztModel};
		\node(zrun) [used, right = .7cm of z3] {\ztRunner};
		\node(nd) [dev, below = of call, yshift =.1cm, minimum width=2.2cm] {\detChecker};
		\node(build) [dev, right = of nd] {\formulaBuilder};
\node(gri) [dev, right = of vald, xshift = -0.2cm, yshift = -1cm] {\trGrinder};
		\node(acons) [dev, below = of nd, minimum width=2.2cm] {\aConsistency};
		\node(an) [dev, right = of acons, yshift = -.7cm] {\ztModelAnalyzer};
		\node(verdict) [output, right = of an, xshift =-.5cm] {\analysisVerdict};
		\node(vfms) [output, above = of zrun] {\fsmImage};

\path[dashed] (txt) -- (vald);
		\path[dashed] (ggen.north) |- (vfms.west);
		
		\path[dashed] (an) -- (verdict);
		
		\node(dummy)[right = .3cm of vald]{};
		\path[-, dashed] (vald) -- (dummy.center);
		\path[dashed] (dummy.center) |- (ggen);
		\path[dashed] (dummy.center) |- (gri);
		
		\node(dummy2)[right = .3cm of gri]{1};
		
		\path[-] (gri) -- (dummy2.center);
		\path (dummy2.center) |- (call);
		\path (dummy2.center)  -- (nd);
		\path (dummy2.center) |- (acons);
		
		\node(dummy21c)[above = .3cm of dummy2]{\mkNode{1}};
		\node(dummy21d)[below = .4cm  of dummy21c]{\mkNode{2}};
		\node(dummy21a)[below = .3cm of dummy2]{\mkNode{3}};
		
		\node(dummy3)[right = .3cm of nd]{};
		\path [dashed] (dummy3.center) -- (build);
		\path[-, dashed] (call.east) -| (dummy3.center);
		\path[-, dashed] (nd.east) -- (dummy3.center);
		\path[-, dashed] (acons.east) -| (dummy3.center);
		
		\path[dashed] (build) -- (z3);
		\path[dashed] (z3) -- (zrun);
		\node(dummy4)[right = .3cm of build,yshift = -.4cm]{};
		\path[-] (zrun) |- (dummy4.center);, dashed
		\path (dummy4.center) -| (an.north);
		
\begin{pgfonlayer}{background}
			\node[fit = (call) (nd) (acons),
			fill=green!70,
			opacity=.7,
			rounded corners,
			minimum width =2.8cm,
			minimum height = 3cm
			] {};
		\end{pgfonlayer}
		\begin{pgfonlayer}{background}
			\node[xshift =.35cm, fit = (dummy) (call) (acons) (build),
			fill=orange,
			opacity=.5,
			rounded corners,
			minimum width = 8.5cm,
			minimum height = 3.2cm,
			label={[xshift=-3.5cm,yshift=-1cm, black] above:{\huge$\circlearrowleft$}}
			] {};
		\end{pgfonlayer}
		\begin{pgfonlayer}{background}
			\node[fit = (vald) (call) (z3) (an.south east) (dummy4),
			fill=yellow!30,
			opacity=.5,
			rounded corners,
			minimum width = 11.5cm,
			minimum height = 4.3cm,
] {};
		\end{pgfonlayer}
	\end{tikzpicture}
	\caption{\label{fig:arch}The architecture of \thetool}
\end{figure}
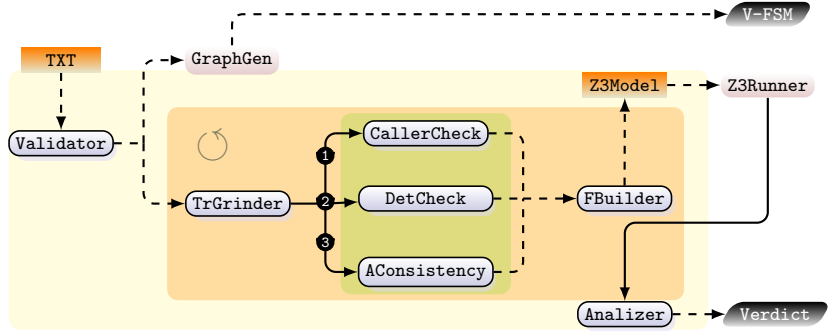

\cref{fig:arch} represents the architecture of \thetool which, for
convenience, is compartmentalised into two principal modules:
\modelname parsing and visualisation (yellow box) and \thetool's core
(orange box).
The latter module implements well-formedness check (green box).
Solid arrows represent calls between components while dashed arrows
data IO.

The flow starts \validator performing basic syntactic checks on a
textual representation\footnote{Our DSL is immaterial here; it is described in the accompanying
  artefact submission.}
of \modelnames and transforming the input in a format that simplifies
the analysis of the following phases.
Specifically, the output of \validator can be passed ($i$) to
\graphGen, a component yielding a visual representation of
\modelnames (\fsmImage\ output) and ($ii$) to the \quo{transitions
  Grinder} \trGrinder component (orange box) for well-formedness
checking.

The component \trGrinder relays each transition of the \modelname in
input to the components in the green box that perform the verification
of well-formedness according to \cref{sec:model}; more precisely:
\begin{itemize}
\item \callerChecker\ (arrow \mkNode{1}) that \toUpdate{returns a boolean}
  which is true if, and only if, the \modelname is closed and strongly
  empty-role free;
\item \detChecker\ (arrow \mkNode{2}) that builds a Z3 formula which
  is true if, and only if, the state
	 is strongly deterministic;
\item \aConsistency\ (arrow \mkNode{3}) to generate a Z3 formula which
  holds if, and only if, the transtion is consistent. 
\end{itemize}

The component \formulaBuilder computes the conjunction of the output of the 
components above, yelding a \ztModel, which is then executed by the \ztRunner.

The verification process ends with the \ztModelAnalyzer\ component
that diagnoses the output of Z3 and produces a \analysisVerdict\ which
reports (if any) the violations of well-formedness of the \modelname in
input.

\subsection{Implementation}\label{sec:impl}
We now give some implementation details on the main features of
\thetool; we first consider each component of \thetool's architecture.

The \validator processes the input which essentially lists transitions
of a \modelname expressed in the format of our DSL.
For instance, the transition to make offers of \cref{ex:smp} is
rendered in our DSL as
\begin{verbatim}
S0 {_offer > 0} b:B > c.makeOffer(int _offer) {offer := _offer} S1 
\end{verbatim}
Basically, \validator reads each transition in the file and extract
participants, actions, states, preconditions, assignments and input
parameters of the action.
To inspect the \modelname in input \thetool relies on \graphGen\footnote{ \toUpdate{\graphGen is a wrap component that uses GraphStream \cite{graphstream} to generate the visual FSM (\fsmImage).}} which creates a visual representation of graphs.

Component \trGrinder transforms the \modelname obtained by \validator
in an internal format suitable for the analysis.
Next, \trGrinder iterates on the transitions, invokes different
checker component by supplying them with the necessary data.

The first component invoked by \trGrinder is \callerChecker which
takes in input a transition $\transT$ and the (internal representation
of the) \modelname.
If the caller of $\transT$ is of the form $\partyP$ or $\any$,
\callerChecker retrieves all acyclic paths\footnote{Crucially, the
  internal format produced by \trGrinder is instrumental for
  extracting acyclic paths using the \txt{networkx}
  library~\cite{networkx}.
} that, from the initial state, lead to
$\transT$'s source state, and then checks that every such path
contains $\new[\partyP][\roleR]$ or $\any$ for some $\roleR$ (if the
caller was $\partyP$), or contains $\new[\partyP][\roleR]$ (if the
caller was $\any$).
As soon as a path violates that condition, \callerChecker halts
returning $\mathsf{False}$ otherwise $\mathsf{True}$ is returned.
To avoid checking again a same path, the formula is saved and just
retrieved when transitions with same same source and caller as
$\transT$ are considered.

The component \detChecker takes as inputs a transition $\transT$ and
the list of transitions with source the target $\transT$.
The list is partitioned by grouping transitions with the same function
name and callers not related by $\partyConfl$ (cf. \cref{sec:model}).
For non-singleton partitions, \detChecker builds a Z3 formula which
is true if, and only if, whenever the guard of a transition is true
the others are false.
Let $\mathcal{T}$ be the set of all transitions and $\mathcal{F}$ be
the partition of $\mathcal{T}$ as described above.
The formula returned by \detChecker is the Z3 correspondent of
$ \Phi_{\text{\detChecker}} = \bigwedge_{F \in \mathcal{F}} \Phi(F)$
where, assuming that $\guardG[\transT]$ is the guard of a transition
$\transT$, we set
\[
  \Phi(F) = \bigwedge_{t \in F} \left( \text{$\guardG[t]$} \implies \toUpdate{\bigwedge}_{\substack{t' \in F,\ t' \neq t}} \neg \text{$\guardG[\transT']$} \right)
\]
Double checking is avoided by keeping track of checked states.

Component \aConsistency implements~\cref{def:consistency}.
Using the formula of the formal definition \quo{as is} however would
be inefficient, because of the presence of universal quantification
and many unnecessary variables and equations (those of the form
$\varCX = \old\; \varCX$).
Universal quantification, as usual with SMT solvers, is dealt with by
just removing quantifiers and negating the formula.
The result of the checker will be negated again at the end.
Unnecessary equations are  removed as follows.
Given a transition $\transT$ and a list of outgoing transitions from
its source, \aConsistency scrutinises the pre-conditions and
post-conditions for shared state variables.
When a variable is used in both conditions, \aConsistency rename the
occurrences of the variable in the pre-conditions by adding the
\_$\old$ suffix.
Likewise, the suffix \_$\old$ is added to state variables $\varX$
occurring in the right-hand side of an assignment of the
post-condition if $\varX$ is assigned in the post-conditions.
Subsequently, the assignments in the post-conditions are transformed
in a conjunction of equations representing the state update.
Finally, \aConsistency constructs a Z3 formula which ensures that given
the pre-conditions and post-conditions bounded by input variables, at
least one precondition of the outgoing transitions should be met.

From the outputs of each of the above components, \formulaBuilder
generates a single formula composed of the conjunction of all the
formulae for each transition.
After going through all transitions, \formulaBuilder compiles all the
generated Z3 formulas to build the \ztModel.
After processing all transitions, \formulaBuilder outputs a Python file
containing the set of Z3 formulae, referred in~\cref{fig:arch} as the
\ztModel.
This model includes all the necessary libraries, variable
declarations, and solver configurations to run the model and determine
its satisfiability.

The component \ztRunner takes this Python file, executes it, and forwards the
results to the \ztModelAnalyzer. If the \ztModel is found to be
satisfiable, it indicates that the \modelname is well-formed;
otherwise, it is deemed non-well-formed.
This final output is the \analysisVerdict.

Finally, we remark that \thetool operates in two modes: a
\texttt{non-stop mode}, which builds and evaluates the entire model
(used for our experimental evaluation) and a \texttt{stop mode}, which
halts immediately as soon as a violation of well-formedness if found.

\section{Evaluation}\label{sec:expl}
We evaluate \modelnames expressiveness and \thetool performance 
using two benchmarks. The first consists of the examples from the Azure BC 
workbench~\cite{azure}, showing how the \modelnames (and the current version of \thetool) 
deals with simple, yet realistic, SCs also used in related work 
(\eg~\cite{predicateabstractions}); the second contains randomly generated 
large examples to stress-test \thetool.

These examples exhibit a variety of features that are essential for
the representation of SCs.
We consider a significant range of features in our analysis, including
inter-contracts interactions (\feature{ICI}), joining of new
participants by-invocation (\feature{BI}) or by participant passing
(\feature{PP}), role revocation (\feature{RR}), and the possibility
for a participant to assume multiple roles (\feature{MPR}).
Our aim is to assess to what degree \thetool can model these features, present in illustrative expressive examples in the literature on SCs.
Our findings are outlined in \cref{tab:feature}.\footnote{Commonplace features such as multiple participants and multiple
	roles are present in all the examples and supported by \thetool.
}
\begin{table}[t]
	\centering
	\caption{Features in the Azure benchmark}\label{tab:feature}
	\begin{tabular}{@{}lccccc@{}}	  
		& \feature{ICI} & \feature{BI} & \feature{PP} & \feature{RR} &  \feature{MPR} 
		\\[.5em]
		\textsf{Hello Blockchain} 				       & \na & \ok & \na & \na & \na \\
		\textsf{Bazaar}							       & \ko & \ok & \na & \na & \na \\ 
		\textsf{Ping Pong}						       & \ko & \ok & \na & \na & \na \\
		\textsf{Defective Component Counter}           & \na & \ok & \ok & \na & \na \\
		\textsf{Frequent Flyer Rewards Calculator}     & \na & \ok & \ok & \na & \na \\
		\textsf{Room Thermostat} 				       & \na & \na & \ok & \na & \na \\
		\textsf{Simple Marketplace} 				   & \na & \ok & \na & \ib & \na \\ 
		\textsf{Asset Transfer} 					   & \na & \ok & \ok & \ib & \na \\
		\textsf{Basic Provenance} 				       & \na & \ok & \ok & \ib & \na \\
		\textsf{Refrigerated Transport} 			   & \na & \ok & \ok & \ib & \ib \\
		\textsf{Digital Locker}					       & \na & \ok & \ok & \ib & \ib
	\end{tabular}
	\\
	{\scriptsize Legend
		\begin{minipage}{.9\linewidth}
			\begin{itemize}
				\item[\ok]: feature present in the example and \thetool successfully handles it
				\item[\ko]: feature present but not supported by \thetool
				\item[\na]: feature not present int to the example
				\item[\ib]: feature present and \thetool supports it with some workarounds
			\end{itemize}
		\end{minipage}
	}\end{table}
Notably, \thetool covers most of the features and the only limitation
is that \thetool does not support yet inter-contracts interactions (\feature{ICI} column).
Notably, we could approximately model the examples with \feature{RR} and \feature{MPR} using some workarounds. In particular, in all the examples featuring \feature{RR} revocation was performed on singleton roles, that is roles that can be played by at most
one participant at a time. Moreover, every revocation is followed by a re-assignment of
the role to a participant. We therefore modelled this situation using a participant
variable for the role. So, $\new[\partyP][\roleR]$ simultaneously assigns role 
to $\partyP$, and revokes $\roleR$ from the previous participant holding it. This has the
drawback that the role cannot be reassigned to a participant formerly holding it.
For \feature{MPR}, in the examples considered the participant with multiple roles was at 
most one. We could therefore add explicit moves for that participant only to emulate
it having two roles.

\medskip
\MMcomm{We now turn our attention to the performance of \thetool, using a b\toUpdate{e}nchmark
	of randomly generated \modelnames.}
More precisely, we evaluated \thetool using a data set of 135 \modelnames\ARcomm[added]{}\footnote{Number fixed to obtain graphs with a few dozens states but increasing number of transitions and qualified participants.
}
randomly generated according to the following process.\footnote{The parameters (fixable as script inputs) were set in order to obtain sufficiently large \modelnames (covering cases with millions of paths) while maintaining the execution time below one hour.
}
\newcommand{\rand}{\ensuremath{\mathsf{rand}}} Let $\rand(i,j)$ be a
random number between number $i$ and $j \geq i$ (we let
$\rand(i) = \rand(1,i)$).
We fix a maximal number of participants $p \in \rand(2,10)$, of
functions $f \in \rand(10,20)$, and of data variables
$v \in \rand(50)$.
For each $s \in \{10, 20, 30\}$ and for each $s \leq t \leq 3s$ such that $t \mod 5 = 0$ we generate five \modelnames, \eMcomm[new]{each
	having $s$ states, by} iterating the following steps until all nodes
are connected and $t$ transitions have been generated:
\begin{itemize}
	\item create $\rand(2,5)$ transitions with source the current state
	and randomly selected target nodes not connected yet (if any,
	otherwise the targets are selected randomly on the whole set of
	nodes);\footnote{We do not spell out the details of the random generation of guards
		and assignments -- they are immaterial for the performance of
		\detChecker, \aConsistency, and \formulaBuilder.
}
	\item for each of the transitions a qualified participants and an
	operation with a number of parameters are randomly selected
	according to $\rand(p)$, $\rand(0,f)$, and $\rand(0,v)$.
\end{itemize}

We measured the performance of \detChecker, \aConsistency, and
\callerChecker by averaging the running time over 
ten 
executions of each generated \modelname.
The experiments were conducted on a Dell XPS 8960, 13th Gen Intel Core (9-13900K) with 32 cores and 32GB RAM running Linux 6.5.0-17-generic (Ubuntu 23.10, 64bit).
The results are reported in the following plots that we now discuss.

\begin{figure}
	\centering
	\includegraphics[width=.47\textwidth]{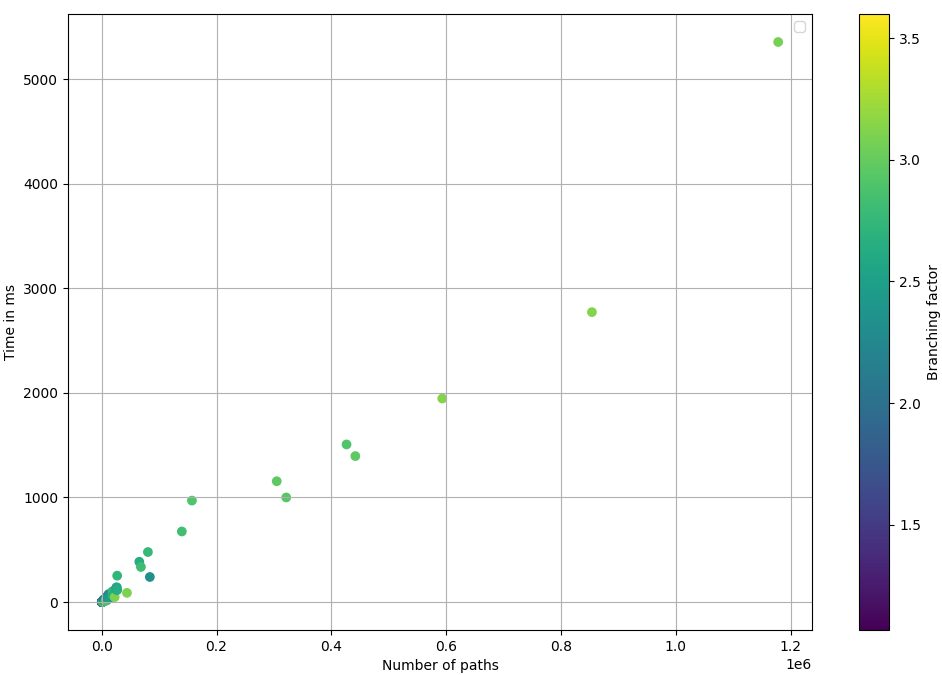}
	\includegraphics[width=.47\textwidth]{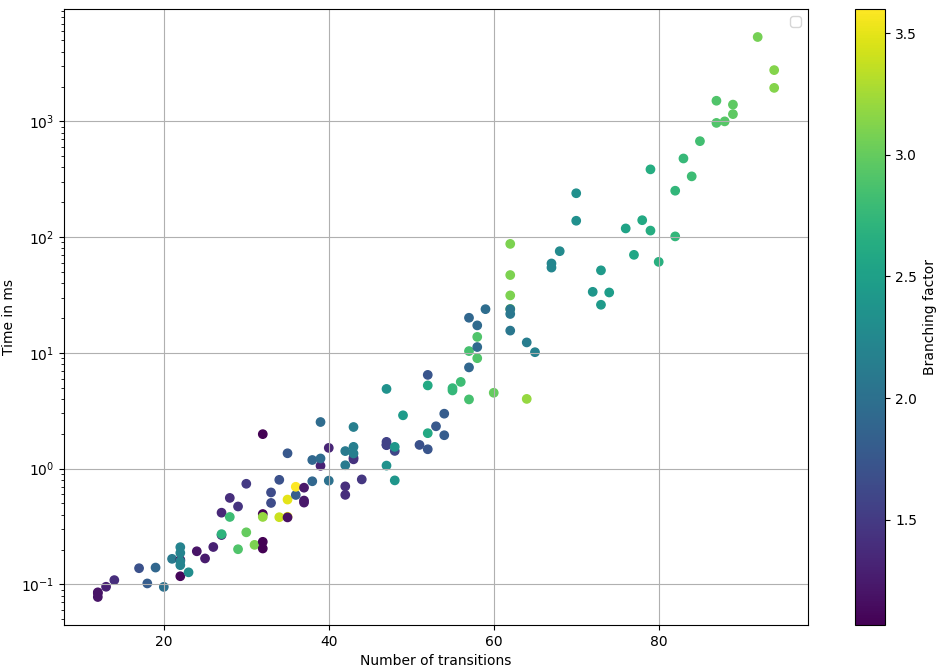}
	\caption{\callerChecker time against of number paths (left) and transitions (right, y-axis in logaritmic scale)}
	\label{fig:evaluation_path_participant_time}
\end{figure}

\begin{figure}
	\centering
	\includegraphics[width=.48\textwidth]{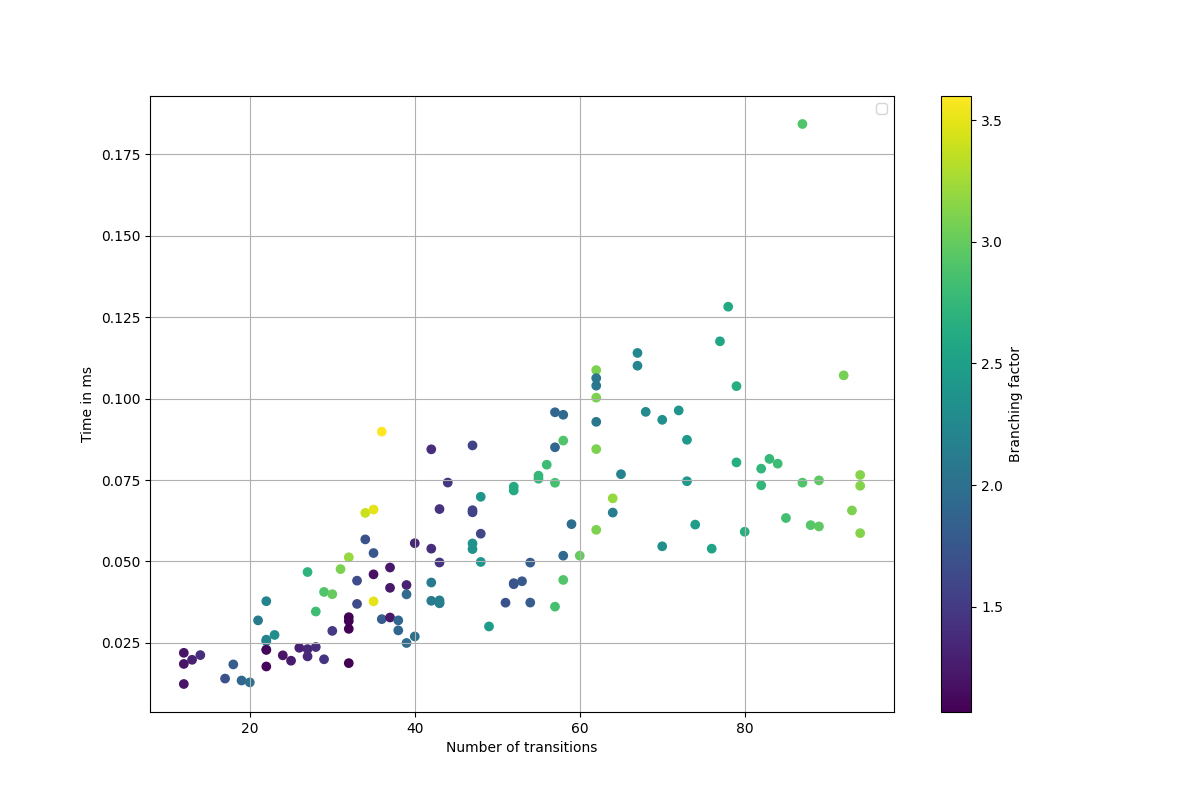}
	\includegraphics[width=.43\textwidth]{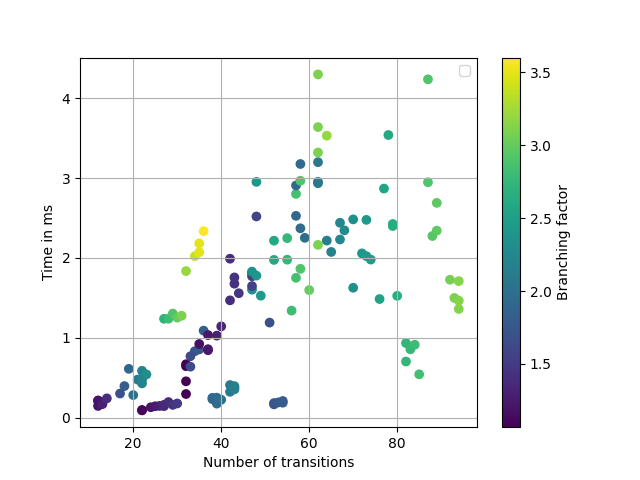}
	\caption{\detChecker (left) \aConsistency (right) time against number of transitions}
	\label{fig:evaluation_states_num_transitions_non_determinism_time_4d}
\end{figure}

The evaluation results presented in~\cref{fig:evaluation_path_participant_time}
\MMcomm{shows the time taken for \callerChecker against the number of paths in the model: 
	as the number of paths increases with the number of transitions of the FSM,
the outcome confirms that times to check closedness and empty-role freeness is exponential in the number of transitions.}
\eMcomm[new]{Interestingly, even in the cases with more that $10^6$ paths,
	\callerChecker terminates the analysis in less that $6$
	seconds.\footnote{To improve the readability of the left plot of
		\cref{fig:evaluation_path_participant_time} we did not include two
		\modelnames whose number of paths was higher more than a factor of
		20 that other instances.
This is not necessary for plot on the right, which in fact
		includes all the generated \modelnames, since there we use a
		logarithmic scale.
}
}

\eMcomm[new]{\cref{fig:evaluation_states_num_transitions_non_determinism_time_4d}
	shows the results of our analysis for \detChecker and \aConsistency.
}
The left plot in
\cref{fig:evaluation_states_num_transitions_non_determinism_time_4d}
hints that the execution time of \detChecker linearly grows with the
number of transitions.
\cref{fig:evaluation_states_num_transitions_non_determinism_time_4d}
right shows that the time to run \aConsistency is too low to allow any conclusion.
Further analysis is require to correlated number of transitions and
time.

\medskip

These plots result from initial investigations of the performances of
the main components of \thetool.\ARcomm[added]{\footnote{Further, more systematic, experiments are needed to lead to broader conclusions.}}
\eMcomm[new]{The complexity of checking well-formedness is dominated by
	\callerChecker, which is exponential in the number of transitions since
	\callerChecker has to check a path property.
However that \thetool shows good performances even for experiments
	with high number of paths
	(cf. \cref{fig:evaluation_path_participant_time}).
}

\section{Related works}\label{sec:rw}
The literature on models of coordination is vast.
We restrict our comparison to tool-supported approaches within three
categories: FSM-based models, formal language models, and
domain-specific languages for SCs.
We compare \modelnames with other coordination models as well as with
approaches specific to SCs.

\medskip

Coordination models of distributed systems based on extensions of FSMs
with (fragments) of first-order logic have appeared in the literature.
Notably data-aware version of BIP and REO have been studied
in~\cite{djab17,qbmhdz19-acta}.
As in \modelnames, data that can be accessed and modified as part of
an interaction in both BIP and REO.
A difference with our model is that interactions can involve more
participants and updates are local to the participants of
interactions.
This also applies to recent models based on asserted communicating
finite-state machines~\cite{pst23,sen23}.

Choreography automata~\cite{blt20coord} and their extension with
assertions~\cite{glsty22} are global specifications for
communicating systems behind
\toolid{Corinne}~\cite{odblt21} and \toolid{CAScr}~\cite{glsty22tool}.
Both these tools are designed to check well-formedness conditions
different than ours (resp. those in~\cite{blt20coord} and
in~\cite{glsty22}) and neither of them supports multiple instances of
roles.
Assertions in \toolid{CAScr} are not guards; they express
rely-guarantee conditions between the sender and the receiver of
interactions.
In the same vein, \toolid{CAT}~\cite{bdft16} is an automata-based tool
for the verification of communication protocols.
Based on \emph{contract automata}~\cite{bt22,bt23,btp20}, \toolid{CAT} is
not data-aware and its contracts purely regard the communication
interface of participants (which are also fixed).

\medskip

Protocol languages that advocates a programming style based on FSMs to
specify SCs are \toolid{FSolidM}
\href{https://github.com/anmavrid/smart-contracts}{framework}~\cite{mavridouL18fsolidm,mavridouEtAl2019verisolid}
and
\textsc{\toolid{SmartScribble}}~\cite{falcaoEtAl2022smartscribble}.
The former relies on model checking CTL formulae to verify safety and
liveness properties (including deadlock-freedom).
The automata have a global state, represented by contract, input, and
output variables, and transitions are guarded by boolean conditions on
these variables.
The tool has been extended to feature code generation and interaction
verification between multiple SCs~\cite{keerthi2023formal}.
This progress marked a substantial improvement in detecting common
vulnerabilities such as re-entrancy attacks and fallback errors.

The interaction patterns that can be programmed with
\textsc{\toolid{SmartScribble}}~\cite{falcaoEtAl2022smartscribble}
correspond to FSMs.
The tool extracts Plutus code\footnote{Plutus is the programming
language to develop SCs to the Cardano BC.} from valid
protocol descriptions, leaving to the developer the task to fill in
the application logic. The automatic generation of code (a feature we
aim to) greatly accelerates developing time, and guarantee
correct-by-construction code (in what concerns the interaction
patterns).

Participants are first-class citizens in \modelnames while
\toolid{FSolidM} encodes them with variables and
\textsc{\toolid{SmartScribble}} identifies participants with roles
which are fixed statically. Also, \textsc{\toolid{SmartScribble}} does
not support assertions.

An application of Event-B to SCs generating automatically Solidity code appeared in~\cite{singh2023formal}. There is no report on the validation of the tool with benchmarks.

A parallel line of research explores the use of BC technology to audit
choreographic programs~\cite{CorradiniM0P0022,CorradiniMMPRT22}.
Roughly, the idea is to generate Solidity contracts from models
expressed as BPMN~\cite{bpmn}, so that the contracts' trace the
execution of their choreography.
Notably,~\cite{CorradiniMultiInstance} is an extension
of~\cite{CorradiniMMPRT22} which allows multiple participants to play
the same role.
This line of work has fairly different goals than ours: its aim is to
exploit BC immutability to record the execution in a secure way.
Our approach instead concerns modelling and verifying distributed
applications coordinated by a FSM, possibly implemented as an SC.

\medskip

The previous tools take a \quo{top-down approach} -- propose an
abstraction to (rigorously) define formal models of computations.
In several cases, SCs code is automatically generated from (correct)
specifications.
The next proposals, \toolid{Obsidian}~\cite{coblenz2020obsidian} and
\toolid{Stipula}~\cite{cra22,crafaEtAl2023stipula,laneve2023stipula-liquidity},
embed the definition of the FSM in the contracts' programming
language.
Both are inspired by \emph{typestates}~\cite{strom1986typestate} and
their use in programming languages~\cite{garcia2014foundations}:
states are explicit entities with a defined API; invocations to an
operation of the API of a state possibly update of the variables
of the program and yield to (possibly) another state.
In this respect, the execution model of both languages is quite similar
to \modelnames.

\toolid{Obsidian} uses typestates and linear types~\cite{lineartypes}
to control \quo{assets} (critical resources of SCs).
Safe, yet flexible, aliasing is ensured with an ownership type
system~\cite{clarkeOSW2013surveyownershiptypes}.
Two case studies established the usability of the language in
\quo{real-world} scenarios.

\toolid{Stipula} focuses on legal contracts and provides a strict
discipline to guarantee \emph{liquidity}: no asset remains frozen
forever.

\medskip

The Azure repository has been used as a benchmark
in~\cite{predicateabstractions} where solidity code is annotated with
assertions.
There the
\href{http://lafhis.dc.uba.ar/dependex/contractor/Welcome.html}{\toolid{Contractor}}
toolkit extracts from the annotate code data-aware abstractions (akin
to FSMs).
Such abstraction can then be validate with respect to user-defined
properties.
The main strength of \toolid{Contractor} is the possibility to
automatically construct sound models, while its main drawback is that
it does not directly support multi object protocols.

\section{Conclusions \& Future works}\label{sec:conc}
This paper proposes \modelnames, a data-aware coordination model for orchestrated computation applicable to the description of multiparty protocols.
The key novelties are:
\begin{enumerate*}
    \item the support for multiple participants, organised by roles, which can dynamically join a protocol;
    \item the use of assertions to describe a protocol state and control how (parametrised) actions change it (in a style akin to Hoare triples);
    \item a notion of well-formed models and a checking algorithm;
    \item a tool for describing systems with \modelnames, visualising them as FSMs, and checking their well-formedness.
\end{enumerate*}

In scope of future work is to define a model-checking approach to
support safety and liveness property analysis.
We also plan generalisations of the model to allow role revocation and
code generation.
An interesting line of work extract \modelnames from actual SC
programs.
A deeper analysis of the performances should also be conducted on more
case studies and possibly refining the random generation of
\modelnames.

\bibliographystyle{splncs}

\end{document}
